\documentclass[a4paper,11pt]{article}
\usepackage{jheppub}
\usepackage[utf8x]{inputenc}
\usepackage[T1]{fontenc}
\usepackage{amsfonts}
\usepackage{amsmath}
\usepackage{amssymb}
\usepackage[titletoc,toc,title]{appendix}
\usepackage{braket}
\usepackage{cancel}
\usepackage[normalem]{ulem}
\usepackage[font={footnotesize}]{caption}
\usepackage{color}
\usepackage{comment}
\usepackage{enumitem}
\usepackage{epsfig}
\usepackage{epstopdf}
\usepackage{float}
\usepackage{graphicx}
\usepackage{hyperref}
\usepackage{cleveref}
\usepackage{mathtools}
\usepackage{ragged2e}
\usepackage{subcaption}
\newcommand{\sech}{\text{sech}}

\hypersetup{
	colorlinks=true,
	citecolor=red,
	filecolor=red,
	linkcolor=blue,
	linktocpage=true,
	urlcolor=blue
}

\begin{document}
	
\title{Islands and dynamics at the interface}

\author{Mir Afrasiar,}
\author{Debarshi Basu,}
\author{Ashish Chandra,}
\author{Vinayak Raj}
\author{and Gautam Sengupta}

\affiliation{
Department of Physics,\\
Indian Institute of Technology,\\ 
Kanpur 208 016, India
}
	
\emailAdd{afrasiar@iitk.ac.in}
\emailAdd{debarshi@iitk.ac.in}
\emailAdd{achandra@iitk.ac.in}
\emailAdd{vraj@iitk.ac.in}
\emailAdd{sengupta@iitk.ac.in}

\abstract{We investigate a family of models described by two holographic CFT$_2$s coupled along a shared interface. The bulk dual geometry consists of two AdS$_3$ spacetimes truncated by a shared Karch-Randall end-of-the-world (EOW) brane. A lower dimensional effective model comprising of JT gravity coupled to two flat CFT$_2$ baths is subsequently realized by considering small fluctuations on the EOW brane and implementing a partial Randall-Sundrum reduction where the transverse fluctuations of the EOW brane are identified as the dilaton field. We compute the generalized entanglement entropy for bipartite states through the island prescription in the effective lower dimensional picture and obtain precise agreement in the limit of large brane tension with the corresponding doubly holographic computations in the bulk geometry. Furthermore, we obtain the corresponding Page curves for the Hawking radiation in this JT braneworld.}

	\maketitle
	
	\flushbottom

\section{Introduction} \label{sec:intro}
In recent years the remarkable progress towards a possible resolution of the black hole information loss paradox in toy models has garnered intense research focus. This development involved the inclusion of  bulk regions termed ``\textit{islands}'', in the entanglement wedge for subsystems in radiation baths at late times \cite{Penington:2019npb, Penington:2019kki, Almheiri:2019hni, Almheiri:2019qdq, Almheiri:2019yqk}. The appearance of these islands ensure that the von Neumann entropy of the Hawking radiation follows the Page curve \cite{Page:1993wv, Page:1993df, Page:2013dx}. The crucial ingredient in this island formalism is the incorporation of  \textit{replica wormhole} saddles dominant at late times, in the gravitational path integral for the R\'enyi entanglement entropy. A more natural way to understand this island formalism was provided through the \textit{doubly holographic} description \cite{Almheiri:2019hni, Sully:2020pza, Rozali:2019day, Chen:2020uac, Chen:2020hmv, Grimaldi:2022suv, Suzuki:2022xwv} in which the radiation baths are described by holographic CFTs dual to a bulk geometry. The island formula then emerges from the standard holographic characterization of the entanglement entropy through the (H)RT prescription in the corresponding higher dimensional bulk geometry.

The above doubly holographic interpretation of the island formula was explored in \cite{Deng:2020ent, Chu:2021gdb} in the context of an extension of the AdS$_3$/BCFT$_2$ \cite{Takayanagi:2011zk, Fujita:2011fp, Rozali:2019day, Sully:2020pza, Kastikainen:2021ybu} duality through the inclusion of additional \textit{defect} conformal matter on the end-of-the-world (EOW) brane. In this defect AdS$_3$/BCFT$_2$ framework the equivalence of the quantum corrected RT formula, termed as the defect extremal surface formula, in the $3d$ bulk geometry with the corresponding island formula in the lower dimensional effective $2d$ description could be demonstrated. This doubly holographic description in the defect AdS$_3$/BCFT$_2$ framework was further investigated for different mixed state entanglement measures \cite{Li:2021dmf, Basu:2022reu, Shao:2022gpg, Lu:2022cgq}.

In relation to the above discussion, the Jackiw-Teitelboim (JT) gravity \cite{Jackiw:1984je, Teitelboim:1983ux} coupled to a radiation bath in two dimensions has proved to be an interesting solvable model to study the application of the island formula \cite{Almheiri:2019qdq, Penington:2019kki}. Recently in \cite{Deng:2022yll}, the authors have realised this setup through a dimensional reduction of a defect AdS$_3$ bulk with small transverse fluctuations on the EOW brane.\footnote{The authors in \cite{Geng:2022slq, Geng:2022tfc} had also obtained the JT gravity from Karch-Randall branes in the context of wedge holography, through a similar prescription.} \footnote{Note that, the authors in \cite{Verheijden:2021yrb, KumarBasak:2021rrx} had investigated a  related prescription to obtain JT black holes through a similar partial dimensional reduction starting from AdS$_3$ geometries.} In particular, they have derived the full JT gravity action through a partial dimensional reduction of the $3d$ bulk wedge sandwiched between a virtual zero tension brane and the finite tension EOW brane, by identifying the transverse fluctuations with the dilaton field in the $2d$ effective description. Usual AdS$_3$/CFT$_2$ prescription has been utilized in the remaining part of the bulk to obtain the bath CFT$_2$ on the asymptotic boundary of the AdS$_3$ geometries. This has provided a $3d$ holographic dual for the JT gravity coupled to a CFT$_2$ bath.

On a separate note, the doubly holographic description of the island formula has been further investigated in \cite{Anous:2022wqh} for an interface CFT (ICFT$_2$) where two CFT$_2$s on half lines with different central charges were considered to be communicating through a common quantum dot. The holographic bulk for such a field theoretic configuration is described by two truncated AdS$_3$ geometries with different length scales, sewed together along the constant tension EOW brane. In such a configuration, the equivalence between the island formula and the holographic entanglement entropy has been illustrated for certain bipartite states. 

In this context, lifting the constraint of the rigidity of the EOW brane in the ICFT$_2$ setup could lead to interesting physics. In the present article this configuration has been investigated where we introduce transverse fluctuations on the EOW brane to obtain the JT gravity through partial dimensional reduction. In the lower dimensional $2d$ effective perspective, this configuration is described by a JT black hole coupled to two CFT$_2$ baths termed CFT$^\text{I}$ and CFT$^\text{II}$. In particular, we have two separate CFT$_2$s in the JT background, interacting through the gravity, whereas they remain decoupled on the remaining half lines with fixed geometry. An alternative description of this configuration involves the consideration of the ICFT$_2$ as a holographic dual of the $3d$ bulk where the interface degrees of freedom may now be interpreted as an SYK quantum dot. Naturally the three perspectives described above constitute a double holographic description for this model of JT gravity coupled to two CFT baths.

We investigate  the entanglement entropy for various bipartite states for this model described by subsystems in the two CFT baths coupled to JT gravity at zero and finite temperatures. Interestingly for our model we encounter certain novel island configurations absent in the earlier analysis with a single bath \cite{Almheiri:2019qdq, Almheiri:2019yqk}. Specifically we demonstrated that there are island contribution from both CFT$^\text{I}$ and CFT$^\text{II}$ even when the subsystem in question involve only bath degrees of freedom from CFT$^\text{II}$ which leads to a modification of the standard island formula involving these {\it induced islands}. Our results for the entanglement entropy obtained through the above modified island formula in the $2d$ effective picture in the large central charge limit, exactly reproduces the $3d$ bulk computations in the doubly holographic perspective.

As a significant consistency check, we perform a replica wormhole computation for one of the configurations considered above to obtain the position of the conical singularity situated at the boundary of the island region. To this end, we employ the well known conformal welding problem \cite{Mumford, Almheiri:2019qdq, Goto:2020wnk} to define a coordinate system consistently spanning the complete hybrid manifold consisting of a gravitational part and two non-gravitating baths. Solving this welding problem reproduces the location of the quantum extremal surface obtained through the extremization of the generalized entropy. It is worth emphasising here that the recovery of the quantum extremal surface from the solution of this welding problem does not assume any holography providing a significant non-trivial consistency check of the island formula for this setup.

The rest of the article is organized as follows. In \cref{sec:review}, we review the basic ingredients of our model, namely, the mechanism of partial dimensional reduction for a defect AdS$_3$ bulk to obtain the JT gravity coupled to a radiation bath, and the salient features of the ICFT$_2$ and its holographic dual. Subsequently, in \cref{sec:JT-EOW} we derive the JT gravity coupled to two radiation baths through a partial dimensional reduction of two truncated AdS$_3$ geometries sewed together along a fluctuating EOW brane. Furthermore, we provide a prescription for the modified island formula in such CFT models. In \cref{sec:Entropy}, we perform the computation for the entanglement entropy for certain configurations in the $2d$ effective description at zero temperature involving extremal JT black holes. Subsequently, in \cref{sec:FiniteT}, the computation of the entanglement entropy for subsystems at finite temperature which involve eternal JT black holes is performed. In \cref{sec:replica}, we provide the replica wormhole computation for a simple configuration considered earlier and show perfect matching with the island result. Finally in \cref{sec:summary} we summarize our results and present conclusions.

\section{Review of earlier literature}\label{sec:review}
\subsection{JT gravity through dimensional reduction}\label{sec:review-JT-reduction}
In this subsection we review the mechanism to obtain JT gravity through the dimensional reduction of an AdS$_3$ geometry truncated by a fluctuating EOW brane \cite{Geng:2022slq, Deng:2022yll, Geng:2022tfc}. For this purpose, consider the defect AdS$_3$/BCFT$_2$ scenario where additional degrees-of-freedom are incorporated at the boundary of the BCFT$_2$ which results in the introduction of defect conformal matter on the EOW brane truncating the AdS$_3$ spacetime. The gravitational action on the dual bulk manifold $\mathcal{N}$ to such a defect BCFT$_2$ defined on the half line $x \geq 0$ is given by \cite{Deng:2020ent, Chu:2021gdb, Shao:2022gpg, Li:2021dmf, Basu:2022reu}
\begin{equation}\label{bulk-action}
	I = \frac{1}{16 \pi G_N} \int_\mathcal{N}\text{d}^3x \sqrt{-g} (R - 2 \Lambda) + \frac{1}{8 \pi G_N} \int_\mathbb{Q} \text{d}^2 y\sqrt{-h} K + I^\text{CFT}_\mathbb{Q} \,,
\end{equation}
where $h_{ab}$ is the induced metric and $K$ is the trace of the extrinsic curvature $K_{ab}$ of the EOW brane denoted as $\mathbb{Q}$. The Neumann boundary condition describing the embedding of the EOW brane $\mathbb{Q}$ with the defect conformal matter is given by
\begin{equation}
	K_{ab} - h_{ab} K = 8 \pi G_N T_{ab} \, ,
\end{equation}
where $T_{ab} = - \frac{2}{\sqrt{-h}}\frac{\delta I^\text{CFT}_\mathbb{Q}}{\delta h^{ab}}$ is the stress energy tensor for the defect CFT$_2$. The authors in \cite{Deng:2022yll} considered the matter action to be of the specific form given by
\begin{equation} \label{matter-action}
	I^\text{CFT}_\mathbb{Q} = - \frac{1}{8 \pi G_N} \int_\mathbb{Q} \text{d}^2 y \sqrt{-h} T \, ,
\end{equation}
where $T$ denotes the brane tension.

A convenient set of coordinates to describe the $3d$ bulk geometry are $(t, \rho, y)$ for which the AdS$_3$ spacetime is foliated by AdS$_2$ slices and the metric is given by
\begin{equation}\label{metric-AdS2-slicing}
	\text{d}s^2 = \text{d} \rho^2 + L^2 \cosh^2 \left( \frac{\rho}{L} \right) \frac{-\text{d} t^2 + \text{d} y^2}{y^2}\,,
\end{equation}
where $L$ is the AdS$_3$ radius. The constant tension $T$ of the brane in these coordinates may then be obtained to be
\begin{equation} \label{tension}
	T = \frac{\tanh \frac{\rho_0}{L}}{L}
\end{equation}
where $\rho_0$ is the location of the brane $\mathbb{Q}$.

\begin{figure}[ht]
	\centering
	\includegraphics[scale=0.8]{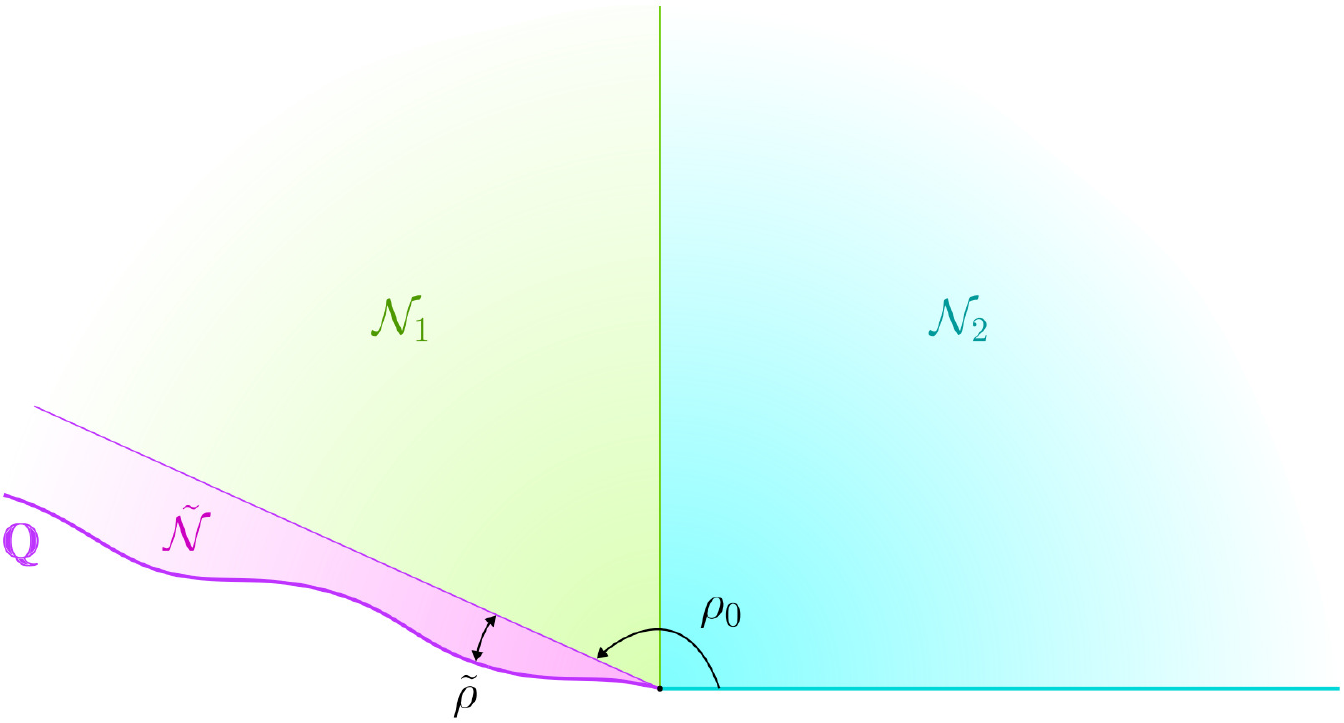}
	\caption{Schematics of the partial dimensional reduction of the AdS/BCFT setup with a fluctuating EOW brane $\mathbb{Q}$. Figure modified from \cite{Deng:2022yll}.}
	\label{fig:JT-red}
\end{figure}

The EOW brane $\mathbb{Q}$ is now made dynamical by introducing a coordinate dependent perturbation of the form \cite{Geng:2022slq, Deng:2022yll}
\begin{equation} \label{fluctuation}
	\rho = \rho_0 + \tilde \rho \,,
\end{equation}
where $\tilde \rho$ is a small fluctuation such that $\frac{\tilde \rho}{\rho_0} \ll 1$. For the specific form of the metric in \cref{metric-AdS2-slicing}, it is possible to integrate out the $\rho$ direction in the bulk for the wedge region $\mathcal{N}_1 + \mathcal{\tilde N}$ as shown in \cref{fig:JT-red}. Dimensional reduction for the region $\mathcal{N}_2$ in the $\rho$ direction will give the original CFT$_2$ on the asymptotic boundary of the AdS$_3$ bulk through the usual AdS/CFT correspondence. 

Now performing this partial dimensional reduction for the bulk gravitational action given in \cref{bulk-action} for the wedge region $\mathcal{N}_1 + \mathcal{\tilde N}$, one may obtain the action for the $2d$ effective theory as follows \cite{Deng:2022yll}
\begin{equation} \label{JT-red-action}
	I_{2d} = \frac{\rho_0}{16 \pi G_N} \int_\mathbb{Q} \text{d}^2 y \sqrt{-g^{(2)}} R^{(2)} + \frac{\rho_0}{16 \pi G_N} \int_\mathbb{Q} \text{d}^2 y \sqrt{-g^{(2)}} \frac{\tilde \rho}{\rho_0} \left( R^{(2)} + \frac{2}{L^2 \cosh^2 \big( \frac{\rho_0}{L} \big) } \right) + \ldots \, ,
\end{equation}
where $g_{ab}^{(2)}$ describes the AdS$_2$ metric with the length scale $L \cosh \big( \frac{\rho_0}{L} \big)$, $R^{(2)}$ is the scalar curvature corresponding to $g_{ab}^{(2)}$ and ellipsis denote $\mathcal{O} \left( {\tilde \rho ^2}/{\rho_0^2} \right)$ terms in the perturbative expansion. The $3d$ Newton's constant $G_N$ is related to that in the $2d$ effective theory $G_N^{(2)}$ as follows \cite{Deng:2020ent, Suzuki:2022xwv}
\begin{equation}
	\frac{1}{G_N^{(2)}} = \frac{\rho_0}{G_N} \, .
\end{equation}
It should be noted here that in \cref{JT-red-action}, the tension of the EOW brane is considered to be the same as in \cref{tension} as the fluctuation \eqref{fluctuation} only changes the tension up to $\mathcal{O} \big( \frac{\tilde \rho ^3}{\rho_0^3} \big)$. Remarkably, \cref{JT-red-action} describes the JT gravity action modulo certain boundary terms\footnote{For the recovery of the complete JT action (including the boundary term), see \cite{Deng:2022yll}.} on identification of $\frac{\tilde \rho}{\rho_0}$ with the dilaton field. This provides us with a mechanism for obtaining JT gravity as a $2d$ effective theory through the partial dimensional reduction of  an AdS$_3$ geometry with a fluctuating EOW brane.


\subsection{Interface CFT} \label{sec:review-ICFT}
In this subsection, we review a class of interface CFT$_2$s (ICFT$_2$s) introduced in \cite{Anous:2022wqh}. Their construction involves two CFT$_2$s defined on half lines coupled through a quantum dot. The bulk dual for such a theory is described by two locally AdS$_3$ geometries separated by a permeable EOW brane. The two CFT$_2$s located at the asymptotic boundary of the AdS$_3$ geometries are labelled as CFT$^\text{I}$ and CFT$^\text{II}$ with central charges $c_\text{I}$ and $c_\text{II}$ respectively, and the corresponding dual bulk locally AdS$_3$ geometries are labelled as AdS$^\text{I}$ and AdS$^\text{II}$ with length scales $L_\text{I}$ and $L_\text{II}$ respectively. In the semi-classical approximation, there is also an intermediate $2d$ effective perspective to describe this configuration, which may be obtained by integrating out bulk degrees of freedom. This results in the brane being characterized by a weakly gravitating system coupled to the original CFT$^\text{I,II}$s. This $2d$ effective perspective will be discussed in detail in \cref{sec:JT-EOW} in the context of the JT gravity on the EOW brane.

The action for the dual bulk geometry describing the above configuration is given by \cite{Anous:2022wqh}
\begin{equation}\label{ICFT-action}
	\begin{aligned}
		I & = \frac{1}{16 \pi G_N} \left[ \int_{\mathcal{B}_\text{I}}\text{d}^3x \sqrt{-g_\text{I}} \left(R_\text{I} + \frac{2}{L_\text{I}^2} \right) + \int_{\mathcal{B}_\text{II}}\text{d}^3x \sqrt{-g_\text{II}} \left(R_\text{II} + \frac{2}{L_\text{II}^2} \right) \right] \\
		&~ + \frac{1}{8 \pi G_N} \left[ \int_\Sigma \text{d}^2 y \sqrt{-h} ( K_\text{I} - K_\text{II} )  -2 T \int_\Sigma \text{d}^2 y \sqrt{-h} \right] \,,
	\end{aligned}
\end{equation}
where $h_{ab}$ is the induced metric and $T$ is the tension of the EOW brane $\Sigma$. The relative minus sign between the two extrinsic curvatures $K_\text{I,II}$ is due to the fact that the outward normal is always taken to be pointing from the AdS$^\text{I}$ to the AdS$^\text{II}$ geometry. The properties of the EOW brane is fixed by requiring it to satisfy certain junction conditions. The first of these demands that the induced metric $h_{ab}$ on the brane be the same as viewed from either of the two AdS$_3^\text{I,II}$ geometries. The second is the Israel junction condition for the brane with the two AdS$_3^\text{I,II}$ geometries on either side which may be expressed as \cite{Anous:2022wqh}
\begin{equation} \label{Israel-ICFT}
	\big( K_{\text{I},ab} - K_{\text{II},ab} \big) - h_{ab} \big( K_{\text{I}} - K_{\text{II}} \big) = -T \, h_{ab} \, .
\end{equation}

Solving these junction conditions will require us to specify the coordinate system describing the $3d$ geometry. To this end, the AdS$_2$ foliation of the AdS$_3$ geometry is chosen again on each patch of the spacetime $\mathcal{B}_\text{I,II}$ as follows\footnote{Here $\rho$ is a hyperbolic angular coordinate which can be related to the usual angle $\chi$ as follows
	\begin{equation}\label{rho_psi}
		\tanh\left(\frac{\rho_k}{L_k}\right)\equiv \sin\chi^{}_k~.
	\end{equation}
	In the rest of the article, the location of the brane at $\rho_k^0$ in this coordinate is represented by $\chi^{}_k = \psi_k$.}
\begin{equation}\label{metric-ICFT-AdS2-slicing}
	\begin{aligned}
		\text{d}s_{\mathcal{B}_{k}}^2 &= \text{d} \rho_{k}^2 + L_{k}^2 \cosh^2 \left( \frac{\rho_{k}}{L_{k}} \right) \tilde{h}_{ab} \, \text{d} y^a  \text{d} y^b \\
		&\equiv \text{d} \rho_{k}^2 + L_{k}^2 \cosh^2 \left( \frac{\rho_{k}}{L_{k}} \right) \frac{-\text{d} t_{k}^2 + \text{d} y_{k}^2}{y_{k}^2} \, , \quad \quad k =  \text{I,II}  \, .
	\end{aligned}
\end{equation}
Here $\tilde{h}_{ab}$ describes the usual Poincar\'e AdS$_2$ metric with unit radius. In these coordinates, the EOW brane is considered to be located at $\rho_k = \rho_{k}^0$ for $k = \text{I,II} $. The first junction condition thus implies the identification of $y_k$ and $t_k$ for both the coordinate patches. Additionally it also enforces the two AdS$_2$ radii to be the same i.e., 
\begin{equation}\label{JC2}
	L_\text{I}^2 \cosh^2 \left( \frac{\rho_{\text{I}}^0}{L_\text{I}} \right) = L_\text{II}^2 \cosh^2 \left( \frac{\rho_{\text{II}}^0}{L_\text{II}} \right) \, .
\end{equation}
The solution to the second junction condition \eqref{Israel-ICFT} fixes the position of the EOW brane as follows \cite{Anous:2022wqh}
\begin{equation}\label{BraneTension}
	\tanh \left( \frac{\rho^0_\text{I}}{L_\text{I}} \right) = \frac{L_\text{I}}{2 T} \left( T^2 + \frac{1}{L_\text{I}^2} - \frac{1}{L_\text{II}^2} \right) \, , \qquad \tanh \left( \frac{\rho^0_\text{II}}{L_\text{II}} \right) = \frac{L_\text{II}}{2 T} \left( T^2 - \frac{1}{L_\text{I}^2} + \frac{1}{L_\text{II}^2} \right) \, .
\end{equation}
Notice from the above that the tension $T$ of the brane has an upper as well as a lower bound. In the large tension limit described by 
\begin{equation}
	T \to T_\text{max} = \frac{1}{L_\text{I}} + \frac{1}{L_\text{II}} \, ,
\end{equation}
the EOW brane approaches the extended asymptotic boundary of both the AdS$_k$ patches. In this limit, integrating out the bulk degrees of freedom on either side results in the two CFT$_2$s interacting through the weakly gravitating brane. This is the intermediate $2d$ effective scenario mentioned earlier which will be discussed in detail in the following section in the context of the JT gravity on the EOW brane.

\section{Realising JT gravity at the interface of two spacetimes}\label{sec:JT-EOW}
\begin{figure}[ht]
	\centering
	\includegraphics[scale=0.7]{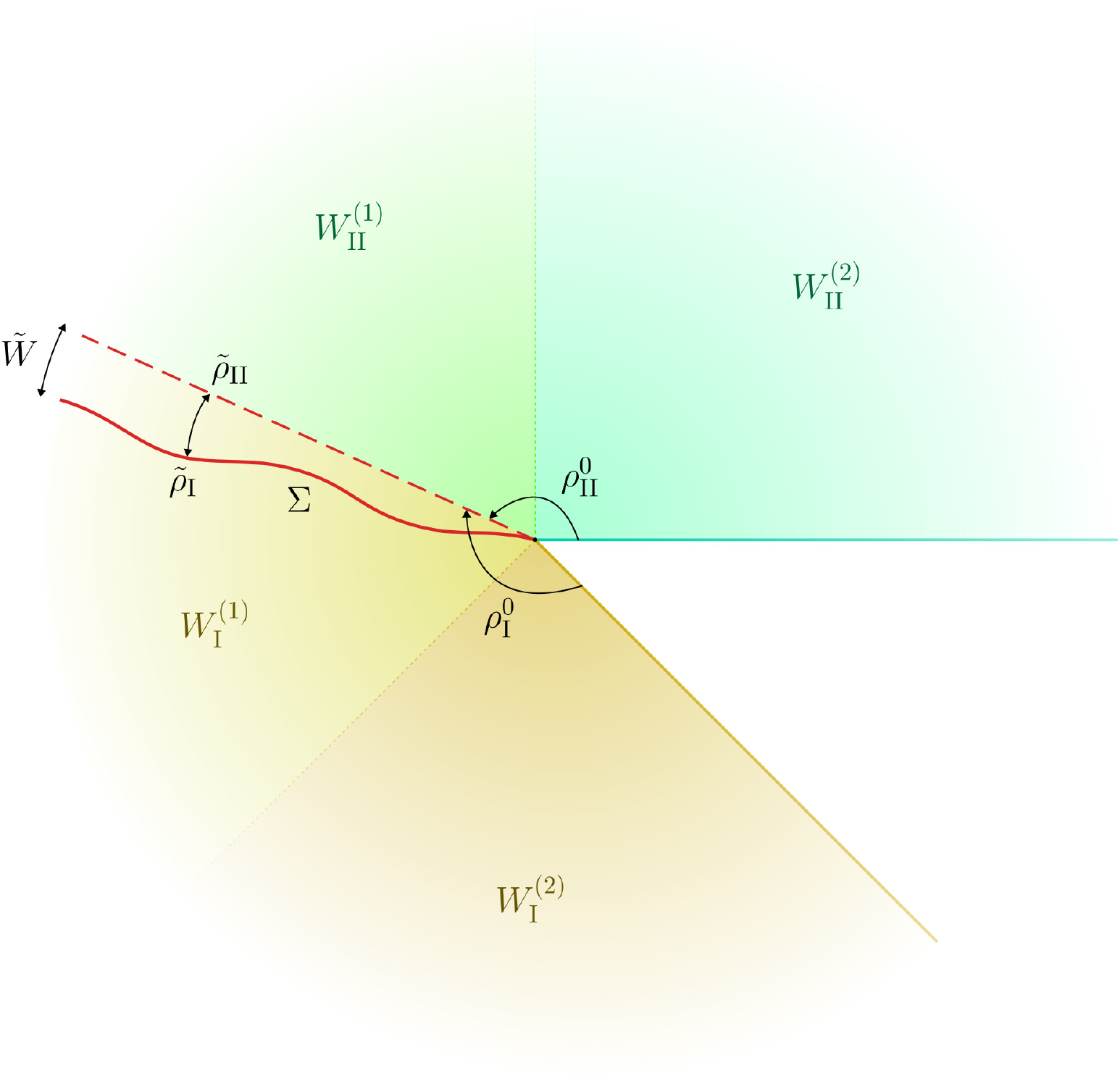}
	\caption{Schematics of the partial dimensional reduction of the AdS/ICFT setup with a fluctuating EOW brane $\Sigma$.}
	\label{fig:IJT-partialdim}
\end{figure}

In this section, we employ a combination of a partial Randall-Sundrum reduction and the usual AdS/CFT correspondence \cite{Deng:2022yll, Deng:2020ent, Chu:2021gdb, Shao:2022gpg, Li:2021dmf, Basu:2022reu} to the AdS/ICFT setup described in the preceding subsection, while allowing for small transverse fluctuations of the EOW brane $\Sigma$. This procedure results in a two dimensional effective theory comprising of the JT gravity on the EOW brane $\Sigma$ coupled to two non-gravitating bath CFT$_2$s. The gravity theory on the brane is obtained by integrating out the bulk AdS$_3$ geometry near the brane and may be thought of as the ``bulk dual'' of the interface degrees of freedom. On introducing the transverse fluctuations the locations of the EOW brane is described as follows 
\begin{align}
	\Sigma~:~&\rho_{\text{I}}=\rho^0_{\text{I}}-\tilde{\rho}_{\text{I}}(y)\notag\\
	&\rho_{\text{II}}=\rho^0_{\text{II}}+\tilde{\rho}_{\text{II}}(y)\,.
\end{align}
The schematics of the setup is depicted in \cref{fig:IJT-partialdim}. In the above equation, $\tilde{\rho}_{k}(y)\ll \rho^0_k$ are the small transverse fluctuations away from the brane angle $\rho^0_k$. Note that the fluctuation modes are functions of the braneworld coordinates $y$ and are treated as fields on the braneworld, as described in \cite{Geng:2022tfc,Deng:2022yll}. As depicted in \cref{fig:IJT-partialdim}, we may divide the two AdS$_3$ geometries on either side of the EOW brane, into the wedges $W_{k}^{(1)}$ and $W_{k}^{(2)}$. With the fluctuations of the brane turned on, the wedge $W_{\text{II}}^{(1)}$ is extended further to include the small wedge region\footnote{Note that the fluctuations of the EOW brane are completely arbitrary in this setting and may as well excise a portion of the wedge $W_{\text{II}}^{(1)}$ from the AdS$_3^{\text{II}}$ instead.} $\tilde{W}$ which is excised out of the wedge region $W_{\text{I}}^{(1)}$. Note that, in this setup the AdS$_3$ spacetimes on either sides of the brane are composed of several wedges as follows
\begin{align}
	\mathcal{B}_{\text{I}}&=W_{\text{I}}^{(1)}+W_{\text{I}}^{(2)}-\tilde{W}\,,\notag\\
	\mathcal{B}_{\text{II}}&=W_{\text{II}}^{(1)}+W_{\text{II}}^{(2)}+\tilde{W}\,.\notag
\end{align}

We now employ the partial dimensional reduction in the wedge regions $W_{\text{I}}^{(1)}-\tilde{W}$ in the AdS$_3^{\text{I}}$ and $W_{\text{II}}^{(1)}+\tilde{W}$ in the AdS$_3^{\text{II}}$ geometries by integrate out the bulk AdS$_3$ degrees of freedom in the $\rho_{\text{I,II}}$ direction(s). On the other hand, in the wedges $W_k^{(2)}$, we utilize the standard AdS$_3$/CFT$_2$ correspondence which leads to flat non-gravitating CFT$_2$s on the half lines stretching out from the interface. It is important to note here that in order to perform a perturbative analysis in $(\tilde{\rho}_{k}/\rho_k^{0})$, it is required to keep $\rho_k^{0}$ large. This restricts us in the large tension regime $T\to T_{\text{max}}$ of the AdS/ICFT setup as advocated in \cite{Anous:2022wqh}. In this limit, the EOW brane $\Sigma$ is pushed towards the asymptotic boundaries of each AdS$_3$ geometry and hence the AdS$_3$ isometries are reminiscent of the conformal transformations on the brane. Therefore, it is natural to expect a gravitational theory coupled to the two bath CFT$_2$s to emerge in the lower dimensional effective description obtained from the partial dimensional reduction. In the following, we investigate the nature of this gravitational theory by explicitly integrating out the bulk AdS$_3$ geometries.

The three-dimensional bulk Ricci scalars are related to the $2d$ Ricci scalar $R^{(2)}$ on the brane $\Sigma$ as follows
\begin{align}
	\sqrt{-g_k}R_k=\sqrt{-g^{(2)}}\left[R^{(2)}-\frac{2\left(3 \cosh^2\left(\frac{\rho_k}{L_k}\right)-1\right)}{L_k^2\cosh^2\left(\frac{\rho_k^0}{L_k}\right)}\right] \, ,
\end{align}
where we have utilized the metric in \cref{metric-ICFT-AdS2-slicing} with 
\begin{align}
	g^{(2)}_{ab} = L_{k}^2 \cosh^2 \left( \frac{\rho_{k}}{L_{k}} \right) \tilde{h}_{ab} \,.
\end{align}	
Integrating the $3d$ bulk Einstein-Hilbert actions (cf. \cref{ICFT-action}) of the AdS$_3^{\text{I,II}}$ regions inside the wedges $W_{\text{I}}^{(1)}-\tilde{W}$ and $W_{\text{II}}^{(1)}+\tilde{W}$ leads to
\begin{align}
	&\frac{1}{16\pi G_N}\int_{W_{\text{I}}^{(1)}-\tilde{W}}\text{d}^3x\sqrt{-g_\text{I}} \left(R_\text{I} + \frac{2}{L_\text{I}^2} \right)+\frac{1}{16\pi G_N}\int_{W_{\text{II}}^{(1)}+\tilde{W}}\text{d}^3x\sqrt{-g_\text{II}} \left(R_\text{II} + \frac{2}{L_\text{II}^2} \right)\notag\\
	&=\frac{1}{16\pi G_N}\int_\Sigma \text{d}^2 y\sqrt{-g^{(2)}}\left[\left(\rho^0_{\text{I}}-\tilde{\rho}_{\text{I}}(y)\right)R^{(2)}-\frac{\sinh\left(\frac{2\rho^0_{\text{I}}-2\tilde{\rho}_{\text{I}}(y)}{L_{\text{I}}}\right)}{L_{\text{I}}\cosh^2\left(\frac{\rho_{\text{I}}^0}{L_{\text{I}}}\right)}\right]\notag\\
	&\quad\quad\quad\quad+\frac{1}{16\pi G_N}\int_\Sigma \text{d}^2 y\sqrt{-g^{(2)}}\left[\left(\rho^0_{\text{II}}+\tilde{\rho}_{\text{II}}(y)\right)R^{(2)}-\frac{\sinh\left(\frac{2\rho^0_{\text{II}}+2\tilde{\rho}_{\text{II}}(y)}{L_{\text{II}}}\right)}{L_{\text{II}}\cosh^2\left(\frac{\rho_{\text{II}}^0}{L_{\text{II}}}\right)}\right]\label{EH-reduction} \, .
\end{align}
Next, we focus on the Gibbons-Hawking boundary terms and the tension term in \cref{ICFT-action}. The extrinsic curvatures $K_{\text{I,II}}$ may be computed using the outward normal vector pointing to $\text{I}\to\text{II}$ as follows
\begin{align}
	K_{\text{I},ab}=\frac{1}{L_\text{I}}\tanh\left[\frac{\rho^0_{\text{I}}-\tilde{\rho}_{\text{I}}(y)}{L_{\text{I}}}\right]\,h_{ab}~~,~~K_{\text{II},ab}=-\frac{1}{L_\text{II}}\tanh\left[\frac{\rho^0_{\text{II}}+\tilde{\rho}_{\text{II}}(y)}{L_{\text{II}}}\right]\,h_{ab} \, ,
\end{align}
where $h_{ab}$ is the induced metric on the brane 
\begin{align}
	h_{ab}=L_{k}^2 \cosh^2 \left( \frac{\rho_{k}}{L_{k}} \right)\tilde h_{ab} \, ,
\end{align}
and $\tilde h_{ab}$ is as defined in \cref{metric-ICFT-AdS2-slicing}. We keep the tension of the fluctuating brane constant as given in \cref{BraneTension}, perturbatively in $\tilde{\rho}_k$. This may be interpreted as the tension of the brane remaining intact under small transverse fluctuations. Hence, the Gibbons-Hawking boundary term together with the brane tension term leads to
\begin{align}
	&\frac{1}{8 \pi G_N} \left[ \int_\Sigma \text{d}^2 y \sqrt{-h} ( K_\text{I} - K_\text{II} )  -2 T \int_\Sigma \text{d}^2 y \sqrt{-h} \right]\notag\\
	&=\frac{1}{8 \pi G_N }\int_{\Sigma}\text{d}^2 y\sqrt{-g^{(2)}}\left[\frac{\sinh\left(\frac{2\rho^0_{\text{I}}-2\tilde{\rho}_{\text{I}}(y)}{L_{\text{II}}}\right)}{L_{\text{I}}\cosh^2\left(\frac{\rho_{\text{I}}^0}{L_{\text{I}}}\right)}-\frac{\tanh\left(\frac{\rho_{\text{I}}^0}{L_{\text{I}}}\right)\cosh^2\left(\frac{\rho^0_{\text{I}}-\tilde{\rho}_{\text{I}}(y)}{L_{\text{II}}}\right)}{L_{\text{I}}\cosh^2\left(\frac{\rho_{\text{I}}^0}{L_{\text{I}}}\right)}\right]\notag\\
	&\qquad\qquad+\frac{1}{8 \pi G_N }\int_{\Sigma}\text{d}^2 y\sqrt{-g^{(2)}}\left[\frac{\sinh\left(\frac{2\rho^0_{\text{II}}+2\tilde{\rho}_{\text{II}}(y)}{L_{\text{II}}}\right)}{L_{\text{II}}\cosh^2\left(\frac{\rho_{\text{II}}^0}{L_{\text{II}}}\right)}-\frac{\tanh\left(\frac{\rho_{\text{II}}^0}{L_{\text{II}}}\right)\cosh^2\left(\frac{\rho^0_{\text{II}}+\tilde{\rho}_{\text{II}}(y)}{L_{\text{II}}}\right)}{L_{\text{II}}\cosh^2\left(\frac{\rho_{\text{II}}^0}{L_{\text{II}}}\right)}\right]\label{GH-Tension} \, .
\end{align}
Adding the contributions from \cref{EH-reduction,GH-Tension} and expanding perturbatively in small $\left(\tilde{\rho}_{k}/\rho_{k}^0\right)$ the total bulk action for the lower dimensional effective gravitational theory on the brane $\Sigma$, upon partial dimensional reduction on the wedges $W_{\text{I}}^{(1)}-\tilde{W}$ and $W_{\text{II}}^{(1)}+\tilde{W}$, becomes
\begin{equation}
	\begin{aligned}
		I_\text{total} = \frac{\rho^0_\text{I} + \rho^0_\text{II}}{16 \pi G_N} \int_{\Sigma}\text{d}^2 y \sqrt{-g^{(2)}}\,& R^{(2)} - \frac{1}{16 \pi G_N} \int_{\Sigma}\text{d}^2 y \sqrt{-g^{(2)}}\,\tilde{\rho}_\text{I}(y) \Bigg[ R^{(2)} + \frac{2}{L_\text{I}^2 \cosh^2\left( \frac{\rho_\text{I}^0 }{L_\text{I}}\right)} \Bigg] \\
		&\quad + \frac{1}{16 \pi G_N} \int_{\Sigma}\text{d}^2 y \sqrt{-g^{(2)}}\, \tilde{\rho}_\text{II}(y) \Bigg[ R^{(2)} + \frac{2}{L_\text{II}^2 \cosh^2\left( \frac{\rho_\text{II}^0 }{L_\text{II}}\right)} \Bigg]\,,\\ \label{JT-IC2}
	\end{aligned}
\end{equation}
where we have neglected terms of order $({\tilde{\rho}_{k}}/{\rho_{k}^0})^2$. Utilizing \cref{JC2}, the above action may be rewritten in the instructive form
\begin{align}
	I_\text{total} = \frac{1}{16 \pi G_N^{(2)}} \left[\int_{\Sigma}\text{d}^2 y \sqrt{-g^{(2)}}\, R^{(2)}+\int_{\Sigma}\text{d}^2 y \sqrt{-g^{(2)}}\,\Phi(y)\left(R^{(2)}+\frac{2}{\ell_{\text{eff}}^2}\right)\right]\,,\label{JT-IC1}
\end{align}
where we have defined the two dimensional Newton's constant $G_N^{(2)}$ and the curvature scale $\ell_{\text{eff}}$ on the brane $\Sigma$ as follows
\begin{align}\label{anglerelation}
	\frac{1}{G_N^{(2)}}=\frac{\rho^0_\text{I} + \rho^0_\text{II}}{G_N}~~,~~\ell_{\text{eff}}=L_\text{I} \cosh \left( \frac{\rho_{\text{I}}^0}{L_\text{I}} \right) = L_\text{II} \cosh \left( \frac{\rho_{\text{II}}^0}{L_\text{II}} \right)\,.
\end{align}
Furthermore, in \cref{JT-IC1}, we have identified the dilaton field $\Phi(y)$ on the brane with the fluctuations of the brane angles $\tilde{\rho}_k(y)$ as follows
\begin{align}\label{phi_y}
	\Phi(y)=\frac{\tilde{\rho}_\text{II}(y)-\tilde{\rho}_\text{I}(y)}{\rho^0_\text{I}+\rho^0_\text{II}}\,.
\end{align}
With these identifications, the $2d$ bulk action in \cref{JT-IC1} precisely takes the form of the action for JT gravity modulo certain boundary terms, with the topological part of the dilaton field $\Phi_0$ set equal to unity. Furthermore, variation of the action with respect to the dilaton field $\Phi(y)$ leads to the Ricci scalar as
\begin{align}
	R^{(2)}=-\frac{2}{\ell_{\text{eff}}^2}=-\frac{2}{L_{k}^2 \cosh^2 \left( \frac{\rho_{k}^0}{L_{k}} \right)}~~,~~k=\text{I,II}\label{R(2)}
\end{align}
which correctly conforms to the fact that the brane is situated at a particular AdS$_2$ slice as seen from either of the bulk AdS$_3$ spacetimes. 

\begin{figure}[ht]
	\centering
	\includegraphics[scale=0.8]{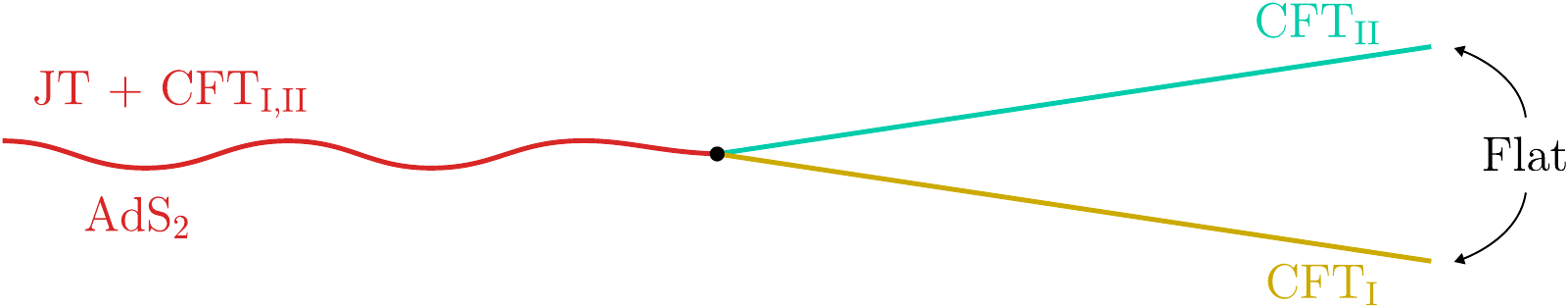}
	\caption{Schematics of the $2d$ effective theory comprised of JT gravity coupled to two flat CFT$_2$ baths.}
	\label{fig:JT+bath}
\end{figure}

At this point we recall that, in the limit of large $\rho_{k}^{0}$ the EOW brane $\Sigma$ is pushed towards the asymptotic boundary of each AdS$_3$ spacetime\footnote{In this limit the tension of the brane is also large, $T\to T_{\text{max}}$, as described in \cite{Anous:2022wqh}.}. As described in \cite{Chen:2020uac,Fallows:2021sge}, in this limit one obtains a non-local action \cite{Skenderis:1999nb} instead of the first term in \cref{JT-IC1} as follows
\begin{align}
	I_{\text{non-local}}=\sum_{k=\text{I} , \text{II}}\frac{L_k}{32\pi G_N}&\int_{\Sigma}\text{d}^2 y \sqrt{-\tilde{h}}\,\left[R^{(2)}-R^{(2)}\log\left(-\frac{L^2_k}{2}R^{(2)}\right)\right]\,.
\end{align}
By introducing two auxiliary scalar fields $\varphi_k~(k=\text{I\,,\,II})$, the above mentioned non-local action may be rewritten in a local form in terms of the usual Polyakov action\footnote{Note that a similar Polyakov action was obtained via covariantization of the induced Liouville action for the gravity theory in the Island/BCFT correspondence described in \cite{Suzuki:2022xwv}.} as discussed in \cite{Anous:2022wqh}
\begin{align}
	I_{\text{Poly}}=\sum_{k=\text{I} , \text{II}}&\frac{L_k}{32\pi G_N}\int_{\Sigma}\text{d}^2 y \sqrt{-\tilde{h}}\left[-\frac{1}{2}\tilde{h}^{ab}\nabla_a\varphi_k\nabla_b\varphi_k+\varphi_kR^{(2)}-\frac{2}{L_k^2}\,e^{-\varphi_k}\right]\,.\label{Polyakov}
\end{align} 
We may interpret the above Polyakov action as two CFT$_2$s\footnote{As explained in \cite{Anous:2022wqh}, the nature of the bulk quantum matter on the brane becomes conformal in the large $\rho_{k}^{0}$ limit.} with central charges $c_{\text{I}}$ and $c_{\text{II}}$ located on the AdS$_2$ brane \cite{Anous:2022wqh}. The JT gravity on the brane is coupled to these CFT$_2$s which are also identical to the two bath CFT$_2$s on the two half lines obtained via the standard AdS$_3$/CFT$_2$ dictionary on the bulk wedges $W_k^{(2)}$. In other words, we have two CFT$_2$s defined on the whole real line. In half of the lines the CFT$_2$s live on a curved AdS$_2$ manifold that is the brane, and are coupled to each other via the JT gravity on this curved manifold. In the other half, the CFT$_2$s live on two flat non-gravitating manifolds and hence are decoupled. The schematics of this $2d$ effective scenario is sketched in \cref{fig:JT+bath}.

To illustrate the emergence of the two CFT$_2$s on the brane via the Polyakov action in \cref{Polyakov}, we note that the zero-dimensional analogue of the transverse area of a codimension two surface $\mathcal{X}$ on the brane is given, for the action \cref{Polyakov}, by \cite{Fallows:2021sge}
\begin{align}
	\mathcal{A}(\mathcal{X})=\frac{\Phi(\mathcal{X})}{4G_N^{(2)}}+\frac{1}{8G_N}\sum_{k=\text{I} , \text{II}}L_k \, \varphi_k(\mathcal{X})\label{AreaPoly}\,.
\end{align}
For the brane $\Sigma$ situated at the AdS$_2$ slice described by \cref{R(2)}, the auxiliary scalar fields $\varphi_k$ may be obtained as \cite{Chen:2020uac}
\begin{align}
	\varphi_k=\log\left[-\frac{2}{L_k^2R^{(2)}}\right]=2\,\log\left[\cosh\left(\frac{\rho_{k}^0}{L_{k}}\right)\right]\,,
\end{align}
and hence the area term in \cref{AreaPoly} is given by
\begin{align}\label{area-term}
	\mathcal{A}(\mathcal{X})=\frac{\Phi(\mathcal{X})}{4G_N^{(2)}}+\frac{c_{\text{I}}}{6}\log\left(\frac{1}{\cos\psi_{\text{I}}}\right)+\frac{c_{\text{II}}}{6}\log\left(\frac{1}{\cos\psi_{\text{II}}}\right)\,,
\end{align}
where we have utilized \cref{rho_psi}.

To conclude we have obtained an effective intermediate braneworld description involving the JT gravity on a dynamical manifold  coupled with two bath CFT$_2$s through a dimensional reduction of a $3d$ bulk which could be understood as a doubly holographic description for the effective $2d$ theory. Recall that this $3d$ bulk has a holographic dual described by an interface CFT where the interface degrees of freedom may be interpreted as an SYK quantum dot.

\subsection{Generalized entropy} \label{sec:S-gen}
Consider a QFT coupled to a gravitational theory on an hybrid manifold $\mathcal{M}=\Sigma \, \cup \mathcal{M}^\text{I}\cup \mathcal{M}^\text{II}$, where $\Sigma$ corresponds to the dynamical EOW brane in the doubly holographic $3d$ description which smoothly joins with the two non-gravitating flat baths\footnote{Note that $\mathcal{M}^\text{I,II}$ forms part of the asymptotic boundary of the $3d$ bulk spacetime, $\partial \mathcal{B}^\text{I,II} \equiv \Sigma \cup \mathcal{M}^\text{I,II}$.} $\mathcal{M}^\text{I,II}$. Transparent boundary conditions are imposed at the common boundary of $\Sigma$ and $\mathcal{M}^\text{I,II}$ such that the quantum matter fields freely propagate across this boundary. The generalized R\'enyi entropy for a subsystem $A$ on this hybrid manifold could be obtained through a path integral on the replicated geometry $\mathcal{M}_n=\Sigma_n\cup \mathcal{M}_n^\text{I}\cup \mathcal{M}_n^\text{II}$ with branch cuts at the endpoints of $A$ as follows
\begin{align}\label{s-gen0}
	(1-n)S^{(n)}_\text{gen}(A)=\log\text{Tr}\boldsymbol {\rho}^n_A=\log\frac{\mathbb{Z}\left[\mathcal{M}_n\right]}{\left(\mathbb{Z}\left[\mathcal{M}_1\right]\right)^n} \, ,
\end{align}
where $\boldsymbol{\rho}_A$ is the reduced density matrix for $A$ in the full quantum theory and $\mathbb{Z} [\mathcal{M}_n]$ corresponds to the partition function of the manifold $\mathcal{M}_n$. Under the semiclassical approximation, the gravitational path integral could be approximated near its saddle point to obtain the partition function on the replicated manifold $\mathcal{M}_n$ as follows
\begin{align} \label{Z-Mn}
	\mathbf{Z}\left[\mathcal{M}_n\right] \approx e^{-\text{I}_\text{grav}\left[\Sigma_n\right]}\,\mathbf{Z}_\text{mat}\left[\mathcal{M}_n\right] \, ,
\end{align}
where 
$\mathbf{Z}_\text{mat}\left[\mathcal{M}_n\right]$ in the matter partition function on the entire replicated hybrid manifold $\mathcal{M}_n$ while $\text{I}_\text{grav} \left[\Sigma_n\right]$ is the classical gravitational action on the dynamical manifold $\Sigma_n$. 

If the replica symmetry for the bulk saddle point configuration in the semiclassical approximation remains intact, the orbifold $\tilde {\mathcal{M}}_n\equiv \mathcal{M}_n/\mathbb{Z}_n$ obtained by quotienting via the replica symmetry $\mathbb{Z}_n$ contains conical defects with deficit angle $\Delta \phi_n = 2 \pi (1-1/n)$ along the replica fixed points in the bulk geometry. This is the so-called \textit{replica wormhole} saddle discussed in the literature \cite{Almheiri:2019hni, Almheiri:2019qdq, Almheiri:2019yqk, Penington:2019kki}. The region enclosed between these conical singularities in the bulk constitute the island $\text{Is} (A)$ for the subsystem $A$. 

In the semiclassical description, the (normalized) matter partition function $\mathbf{Z}_\text{mat}$ computes the effective R\'enyi entropy of the quantum matter fields inside the entanglement wedge of $A \cup \text{Is} (A)$ as follows
\begin{equation}
	\frac{\mathbf{Z}_\text{mat}[\mathcal{M}_n]}{\big(\mathbf{Z}_\text{mat}[\mathcal{M}_1]\big)^n} = \text {e}^{(1-n) \log \text{Tr} \rho_{A \cup \text{Is} (A)}^n } \, ,
\end{equation}
where $\rho_{A \cup \text{Is} (A)}$ is the effective reduced density matrix in the semiclassical description.

\begin{figure}[ht]
	\centering
	\begin{subfigure}[b]{0.45\textwidth}
		\centering
		\includegraphics[scale=0.6]{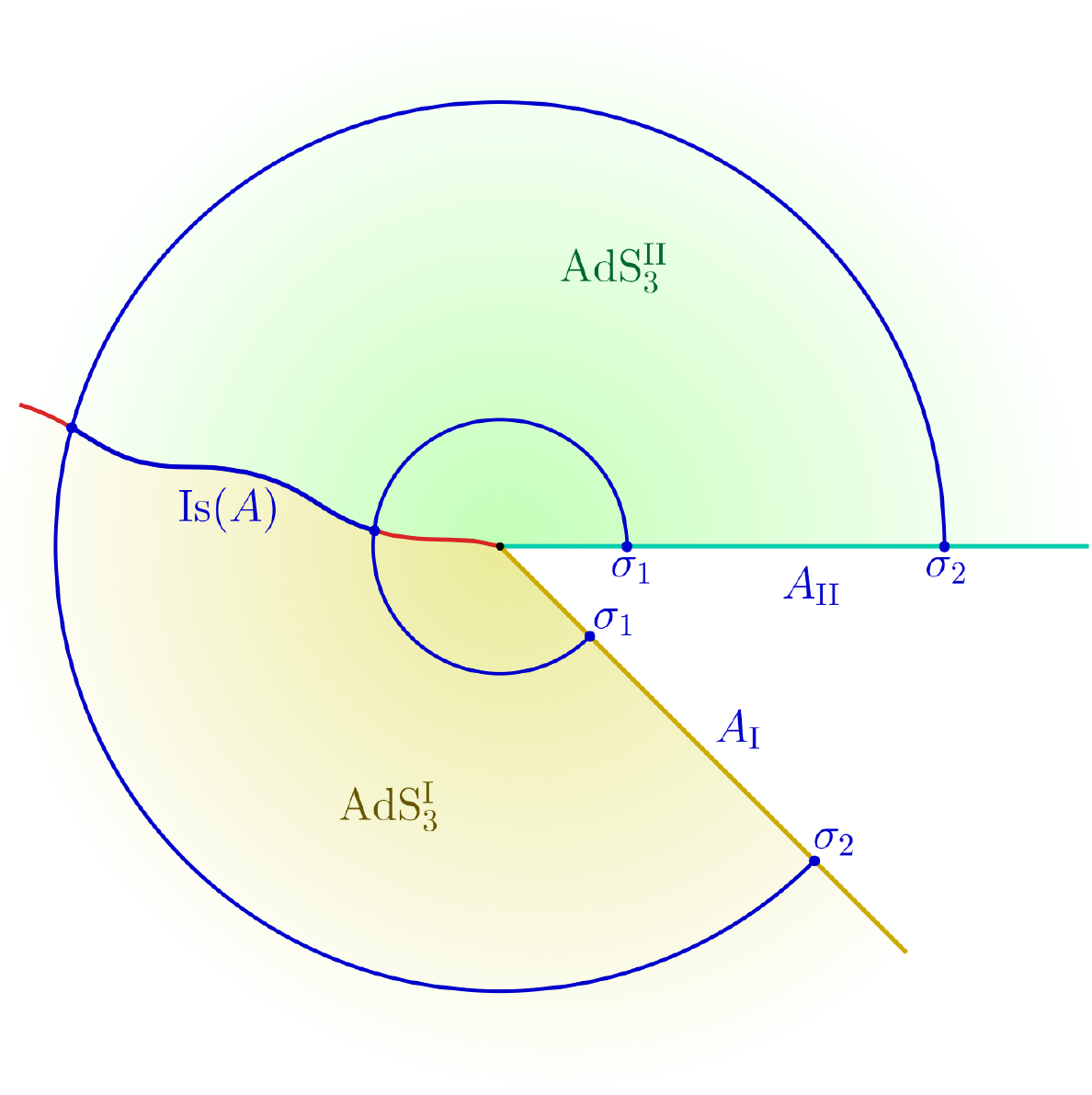}
		\caption{Schematics of the bulk geodesic homologous to the subsystem $A = A_\text{I} \cup A_\text{II}$ described by a finite interval $A_\text{I,II}$ in dual CFT$^\text{I,II}$.}
		\label{fig:finite-single}
	\end{subfigure}
	\hspace{0.8cm}
	\begin{subfigure}[b]{0.45\textwidth}
		\centering
		\includegraphics[scale=0.6]{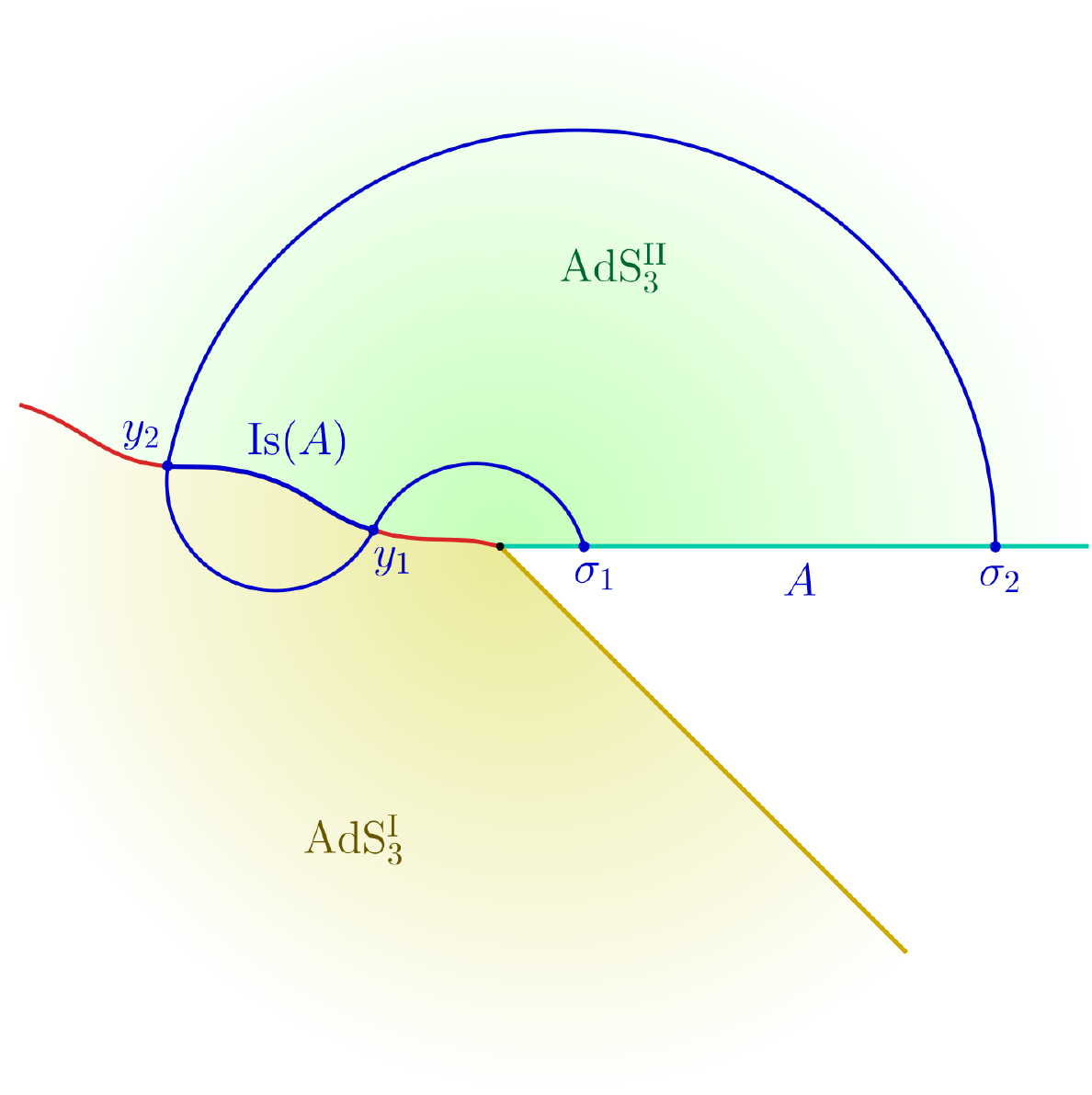}
		\caption{Schematics of the (double-crossing) bulk geodesic homologous to the subsystem $A$ described by a finite interval [$\sigma_1,\sigma_2$] in dual CFT$^\text{II}$.}
		\label{fig:finite-double}
	\end{subfigure}
	\caption{Two independent mechanisms for the origin of the island region $\text{Is}(A)$ in ICFTs.}\label{fig:2islands}
\end{figure}

Unlike the earlier works where JT gravity was coupled to a single radiation bath, in the current scenario, the presence of two baths modifies the structure of the dominant replica wormhole saddle to provide two independent mechanisms for the origin of the island region in the semiclassical description:
\begin{itemize}
	\item For a subsystem $A=A^{\text{I}}\cup A^\text{II}$ with $A^{\text{I,II}}\subset \mathcal{M}^\text{I,II}$ in the radiation baths, both $A^{\text{I}}$ and $A^{\text{II}}$ are responsible for the conical singularities appearing in the gravitating manifold $\Sigma$. In this situation, the corresponding island region $\text{Is}(A)$ manifested in $\Sigma$ depends upon the degrees of freedom for both the CFT baths. In other words, if we denote the islands corresponding to the individual baths as $\text{Is}^{\text{I,II}}(A)$, for the present configuration we have $\text{Is}^{\text{I}}(A)=\text{Is}^{\text{II}}(A)\equiv\text{Is}(A)$ and the density matrix in the effective theory factorizes in the following way
	\begin{align}
		\rho_{A \cup \text{Is} (A)}\sim \rho_{A^\text{I} \cup \text{Is}(A)}\otimes \rho_{A^\text{II} \cup \text{Is} (A)}\,.\label{conventional-rho}
	\end{align}
	This could also be understood through the doubly holographic formalism where we have gravitational regions on either sides of the fluctuating EOW brane. Recall that in the doubly holographic description, the island region in this scenario is described by the region on the EOW brane between the two RTs crossing from AdS$^\text{I}$ to AdS$^\text{II}$. For the present configuration the bulk RT surface homologous to the subsystem $A$ is composed of two geodesics connecting the endpoints of $A^{\text{I}}$ and $A^{\text{II}}$, each of which crosses the EOW brane only once as depicted in \cref{fig:finite-single}. This corresponds to the conventional origin of the island region as described in \cite{Almheiri:2019qdq, Almheiri:2019yqk}.

	\item On the other hand, consider a subsystem $A$ residing entirely in the bath $\mathcal{M}^\text{II}$. If the central charge of the CFT$^\text{II}$ is larger than that of the CFT$^\text{I}$, depending upon the size of the subsystem $A$, conical singularities in the gravitating region $\Sigma$ may appear solely due to the presence of $A$ in the bath $\mathcal{M}^\text{II}$. Since the bulk region $\Sigma$ is common between the bath CFTs, the CFT$^\text{I}$ degrees of freedom present in $\Sigma$ sense the same conical singularities and conceive an \textit{induced island} which we denote by $\text{Is}^{(\text{I} \backslash \text{II})}(A)$ to indicate that we obtain an island region in CFT$^\text{I}$ given some subsystem $A$ in CFT$^\text{II}$. In this case, the density matrix in the effective theory reduces to
	\begin{align}
		\rho_{A \cup \text{Is} (A)}\sim \rho_{\text{Is}^{(\text{I} \backslash \text{II})}(A)}\otimes \rho_{A^\text{II} \cup \text{Is} (A)}\,.\label{novel-rho}
	\end{align}
	From the doubly holographic perspective, this corresponds to a double-crossing geodesic where the minimal curve penetrates into AdS$^\text{I}$ and returns to AdS$^\text{II}$ in order to satisfy the homology condition which is depicted in \cref{fig:finite-double}. Note that such an island region is a novelty of the present model where a gravitational theory is coupled to two flat baths.
\end{itemize}
Assuming that the backreactions from the conical defects are small, the replica wormhole saddle are still solutions to the Einstein's equations. In such case, the classical gravitational action $\text{I}_\text{grav} [\Sigma_n]$ in \cref{Z-Mn}, for the replica wormhole saddle \cite{Almheiri:2019qdq, Dong:2020uxp} may be expressed for $n \sim 1$ as
\begin{equation} \label{Igrav}
	\text{I}_\text{grav} \big[\Sigma_n\big] \approx n \, \text{I}_\text{grav} \big[\tilde {\Sigma}_1\big] + \frac{n-1}{4 G_N^{(2)}} \mathcal{A} \big(\partial \text{Is}(A)\big) \, ,
\end{equation}
where $\tilde {\Sigma}_n \equiv \Sigma_n/\mathbb{Z}_n$ is the orbifold for the replicated bulk geometry $\Sigma_n$ and $\mathcal{A} \big(\partial \text{Is}(A)\big)$ is the area of the boundary of the island region $\text{Is}(A)$, namely the quantum extremal surface for the subsystem $A$. Utilizing \cref{s-gen0,Z-Mn,Igrav}, we may now obtain the generalized entropy corresponding to the reduced density matrix $\boldsymbol {\rho}_A$ for the subsystem $A$ as follows
\begin{align}
	S_{\text{gen}}\left(\boldsymbol {\rho}_A\right)&=\frac{\mathcal{A} \big(\partial \text{Is}(A)\big)}{4G_N^{(2)}}+S_{\text{eff}}\left(\rho_{A \cup \text{Is} (A)}\right)_{\mathcal{M}^\text{I}\cup\mathcal{M}^\text{II}\cup\Sigma}\notag\\
	&=\frac{\mathcal{A} \big(\partial \text{Is}(A)\big)}{4G_N^{(2)}}+S_{\text{eff}}^{\text{I}}\left(\rho_{A^\text{I} \cup \text{Is}(A)} \right)_{\mathcal{M}^\text{I}\cup\Sigma}+S_{\text{eff}}^{\text{II}}\left(\rho_{A^\text{II} \cup \text{Is} (A)}\right)_{\mathcal{M}^\text{II}\cup\Sigma} \, , \label{conventional-Sgen}
\end{align}
for the conventional island. Note that the subscripts $\mathcal{M}^\text{I,II}\cup\Sigma$ denote that the reduced density matrices $\rho_{A^\text{I,II} \cup \text{Is}(A)}$ in the effective theory have support on corresponding manifolds. On the other hand, for the configuration where we observe the induced island, the generalized entropy modifies to
\begin{align}
	S_{\text{gen}}\left(\boldsymbol {\rho}_A\right)=\frac{\mathcal{A} \big(\partial \text{Is}(A)\big)}{4G_N^{(2)}}+S_{\text{eff}}^{\text{I}}\left(\rho_{\text{Is}^{(\text{I} \backslash \text{II})}(A)}\right)_{\Sigma}+S_{\text{eff}}^{\text{II}}\left(\rho_{A^\text{II} \cup \text{Is} (A)}\right)_{\mathcal{M}^\text{II}\cup\Sigma} \, .\label{novel-Sgen}
\end{align}
Note that the area term in \cref{conventional-Sgen,novel-Sgen} for the generalized entropies are as given in \cref{area-term}.

\section{Islands in extremal JT black holes}\label{sec:Entropy}
In this section, we will compute the entanglement entropies of various subsystems at a zero temperature in the CFT$_2^\text{I}$ and CFT$_2^\text{II}$ baths in the braneworld setup discussed above. In particular, we will compute the entanglement entropy for the corresponding subsystems in the intermediate picture using the island formula. Subsequently, we will substantiate these field theory results from the bulk computation of the RT surfaces corresponding to the subsystem using double holography in the large tension limit in which the gravity on the brane is weakly coupled. 

As described in \cite{Almheiri:2019hni,Almheiri:2019qdq,Almheiri:2019yqk}, the dilaton profiles may be obtained from the equation of motion which arises from the JT action in \cref{JT-IC1} by varying it with respect to the metric for the case of extremal black hole as follows
\begin{align}\label{phi}
	\text{d}s^2=\frac{4\,\text{d}\zeta_k \text{d}\overline{\zeta}_k}{\left(\zeta_k+\overline{\zeta}_k\right)^2}~~,~~~~~~\Phi=\Phi_0-\frac{2\Phi_r}{\zeta_k+\overline{\zeta}_k}\,,
\end{align}
where $\zeta = x + i t_E$ are the planar coordinates and $\Phi_0$ is the topological contribution to the dilaton given in \cref{area-term}.

\subsection{Semi-infinite subsystem}\label{sec:single-crossing}
We consider the case where a subsystem $A$ is comprised of a semi-infinite interval in each bath CFT$_2$s as $A \equiv [\sigma_1,\infty]_\text{I} \cup [\sigma_2,\infty]_\text{II}$. We describe the computation of the entanglement entropy of the subsystem $A$ using the island prescription in the effective $2d$ description discussed in \cref{sec:S-gen}. Later we utilize the Ryu-Takayanagi (RT) prescription \cite{Ryu:2006bv} to compute the entanglement entropy of the corresponding interval in the doubly holographic framework. 


\subsection*{Effective $2d$ description}
For this configuration involving a semi-infinite subsystem, only the conventional island appears. Consider the QES to be located at $-a$ on the EOW brane. Note that both the CFT$_2^\text{I}$ and CFT$_2^\text{II}$ are located on the JT brane, thus as discussed in \cref{sec:S-gen}, the conical singularity at the QES $-a$ is present in both CFT$_2^\text{I}$ and CFT$_2^\text{II}$. As can be inferred from \cref{phi}, the UV cutoff on the JT brane has position dependence as $\epsilon(-a)=a$. Hence, utilizing \cref{conventional-Sgen}, the generalized entanglement entropy for subsystem $A$ may be obtained as\footnote{In the following we will set $4 G_N^{(2)}=1$ for brevity.} 
\begin{equation}\label{Sgen-zero-infinite}
	S_{\text{gen}}=\frac{\Phi_r}{a}+\frac{c_{\text{I}}}{6}\log\left[\frac{1}{\cos\psi_{\text{I}}}\right]+\frac{c_{\text{II}}}{6}\log\left[\frac{1}{\cos\psi_{\text{II}}}\right]+\frac{c_\text{I}}{6} \log \bigg[\frac{\left(\sigma _1+a\right)^2}{\epsilon\,a}\bigg]+\frac{c_\text{II}}{6} \log \bigg[\frac{\left(\sigma _2+a\right)^2}{\epsilon\, a}\bigg]\,,
\end{equation}
where we have used \cref{area-term} for the area of the quantum extremal surface located on the JT brane. The entanglement entropy may now be obtained through the extremization of the above generalized entropy over the position of the island surface. The extremization for arbitrary $\sigma_1$ and $\sigma_2$, however leads to complicated expressions. Thus for simplicity, we assume the symmetric case $\sigma_1=\sigma_2=\sigma$, for which the extremization equation is given by
\begin{equation}
	\partial_a S_\text{gen}=0\quad \Rightarrow  \quad \quad a \left(c_{\text{I}}+c_{\text{II}}\right) (a-\sigma )-6 (a+\sigma ) \Phi_r\ =0 \,.
\end{equation}
Finally, the location of the island region $a^*$ may be obtained from the above quadratic equation as follows
\begin{equation} \label{a-extr-inf-T-0}
	a^*=\frac{\left(c_{\text{I}}+c_{\text{II}}\right) \sigma +6 \Phi_r+\sqrt{\big(\left(c_{\text{I}}+c_{\text{II}}\right) \sigma +6 \Phi_r\big)^2+24 \left(c_{\text{I}}+c_{\text{II}}\right) \sigma  \Phi_r}}{2 \left(c_{\text{I}}+c_{\text{II}}\right)} \,,
\end{equation}
where we have disregarded the unphysical solution of the QES. The fine-grained entropy for the subsystem $A$ may finally be obtained by substituting the above extremal value in \cref{Sgen-zero-infinite}. In order to compare this result with the doubly holographic computation in the following subsection, we need to consider the large tension limit of the JT brane for which the brane angles $\psi_\text{I,II}$ may be expanded as \cite{Anous:2022wqh}
\begin{equation}\label{tensionlimit}
	\begin{aligned}
		\psi_\text{I}=\frac{\pi}{2}-\frac{L_\text{I}}{L_\text{I}+L_\text{II}}\delta\,,  \quad\,\quad   \psi_\text{II}=\frac{\pi}{2}-\frac{L_\text{II}}{L_\text{I}+L_\text{II}}\delta\,, \qquad \quad  \text{with } \delta \rightarrow 0\,,
	\end{aligned}
\end{equation}
where the finite but small $\delta$ describes the deviation of the JT brane from the extended conformal boundary of the AdS$_3^\text{I,II}$.

\begin{figure}[ht]
	\centering
	\includegraphics[scale=0.76]{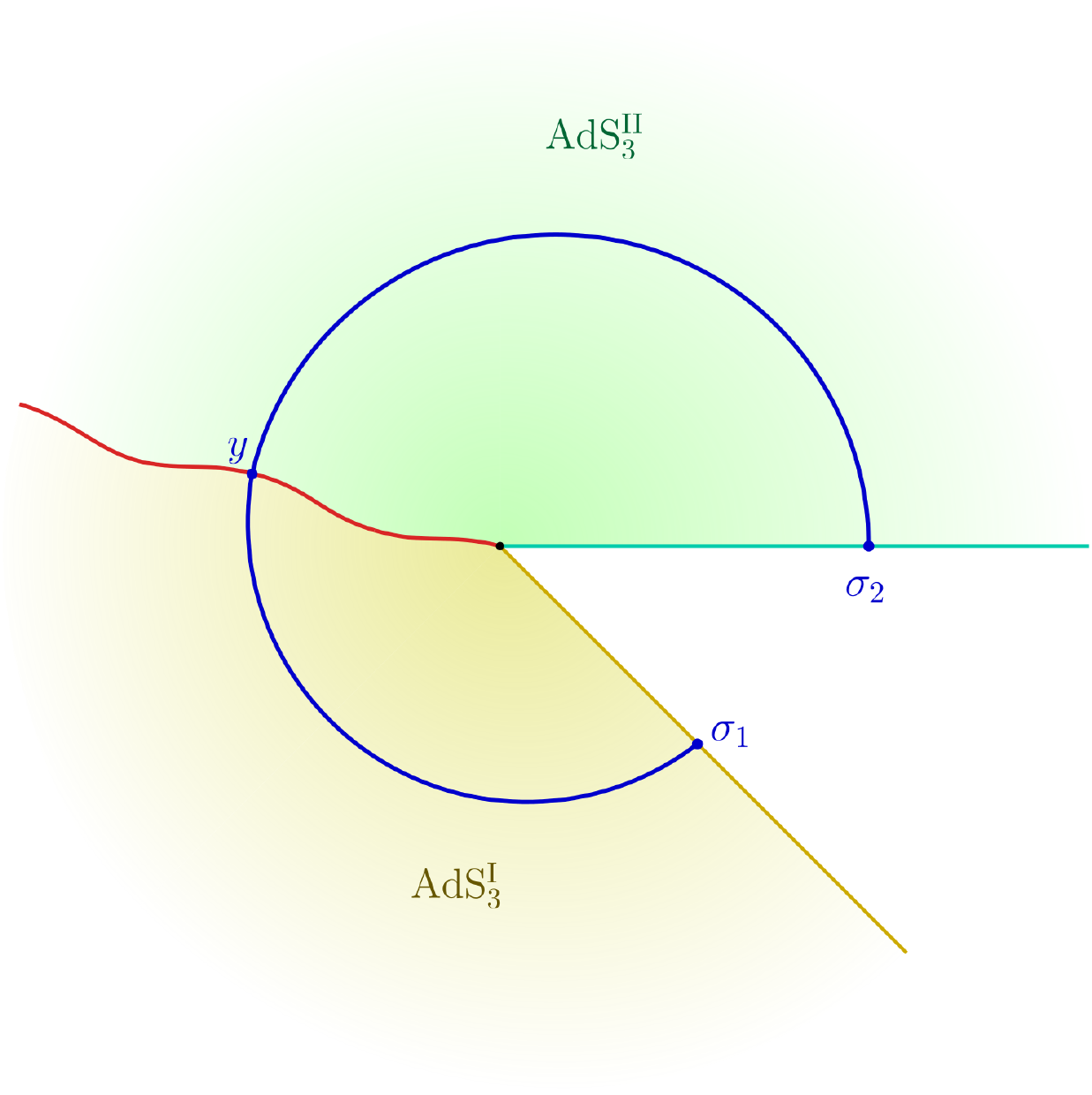}
	\caption{Schematics of the bulk geodesic homologous to the subsystem $A$ described by the union of two semi-infinite intervals in both CFT$^\text{I}$ and CFT$^\text{II}$.}
	\label{fig:single}
\end{figure}

\subsection*{Doubly holographic description}
In this subsection, we substantiate the above island results in the effective field theory from a doubly holographic perspective. To this end, the metric described in \cref{metric-ICFT-AdS2-slicing} may be mapped to the Poincar\'{e} AdS$_3$ geometry through the following coordinate transformations 
\begin{equation}
	z_k= y\cos\chi_k~, \qquad x_k= y \sin\chi_k
	\label{zxToyChi}~.
\end{equation}
The radial direction in the $xz$-plane in the Poincar\'e coordinates is described by the coordinate $y$. At the asymptotic boundary described by $\chi_k=\pm \pi/2$, $y$ now serves as a boundary coordinate. Furthermore, the length of a geodesic between points $(t,x,z)$ and $(t',x',z')$ in the Poincar\'{e} coordinates, is obtained through
\begin{equation}\label{geoleninpoin}
	{d}=L\, \cosh^{-1}\left[\frac{-(t-t')^2+(x-x')^2+z^2+z'^2}{2 z\, z'}\right]~. 
\end{equation}

Note that for the present configuration, the RT surface homologous to subsystem $A$ consists of two semi-circular geodesic segments in each of the AdS$_3^{\text{I,II}}$ geometries which are smoothly joined at the EOW brane as depicted in \cref{fig:single}. As a consequence of the Israel junction condition, we may choose the common point on the EOW brane to be parametrized by a single variable $y$. The total length of the RT surface may then be expressed as
\begin{align}\label{geo_semi1}
	{d}=& L_\text{I}\log\left[\frac{\left(\sigma _1+y \sin \psi _\text{I} \right){}^2+\left( y \cos \psi_\text{I}\right)^2}{\epsilon \, y \cos \psi_ \text{I}}\right] + L_\text{II} \log \left[\frac{\left(\sigma _2+y \sin \psi_\text{II}\right)^2+\left(y \cos \psi _\text{II}\right)^2}{\epsilon \, y \cos \psi _\text{II}}\right]\,,
\end{align}
Subsequently, we perturb the EOW brane by introducing a small fluctuation in the brane angles $\psi _\text{I,II}$ as follows
\begin{align}\label{perturb}
	\psi _{{k}} (y)\to \sin^{-1}\left[\tanh\left( \frac{\rho_{{k}}^0+(-1)^k\tilde{\rho}_k(y)}{L_k}\right)\right]~,
\end{align}
where $\frac{\tilde{\rho}_\text{{I,II}}}{\rho_\text{{I,II}}^0} \ll 1$. Utilizing the above relation, \cref{geo_semi1} reduces to
\begin{align}\label{geo_semi2}
	{d}  &=L_\text{I} \log \left[\frac{ 2 \sigma _1 y \tanh \left(\frac{\rho_\text{I}^0}{L_\text{I}}\right)+\sigma _1^2+y^2}{\epsilon \, y\, \sech \left(\frac{\rho_\text{I}^0}{L_\text{I}}\right)}\right]+L_\text{II} \log \left[\frac{ 2 \sigma _2 y \tanh \left(\frac{\rho_\text{II}^0}{L_\text{II}}\right)+\sigma _2^2+y^2}{\epsilon \, y \, \sech \left(\frac{\rho_\text{II}^0}{L_\text{II}}\right)}\right]\notag\\
	&-\frac{\tilde{\rho}_{\rm I} \left(\sigma _1^2 \tanh \left(\frac{\rho_\text{I}^0}{L_\text{I}}\right)+y^2 \tanh \left(\frac{\rho_\text{I}^0}{L_\text{I}}\right)+2 \sigma _1 y\right)}{2 \sigma _1 y \tanh \left(\frac{\rho_\text{I}^0}{L_\text{I}}\right)+\sigma _1^2+y^2} +\frac{\tilde{\rho}_{\rm II} \left(\sigma _2^2 \tanh \left(\frac{\rho_\text{II}^0}{L_\text{II}}\right)+y^2 \tanh \left(\frac{\rho_\text{II}^0}{L_\text{II}}\right)+2 \sigma _2 y\right)}{2 \sigma _2 y \tanh \left(\frac{\rho_\text{II}^0}{L_\text{II}}\right)+\sigma _2^2+y^2}~.
\end{align}
where we have considered terms up to the first order in $\tilde{\rho}_{\rm I,II}$. Note that, the perturbative parameters $\tilde{\rho}_{\rm I,II}$ are functions of the island location $y$. Therefore, in the perturbative terms of the above expression, we replace $y$ with its zeroth order solution in $\tilde{\rho}_\text{I,II}$ such that the geodesic length in \cref{geo_semi2} contains terms truly upto the first order in $\tilde{\rho}_\text{I,II}$. Subsequently, on identifying the dilaton as given in \cref{phi_y}, the candidate entanglement entropy may be obtained as follows
\begin{align}\label{EE}
	S_{\rm single}(\sigma,y) = \frac{\Phi_r}{y} + \frac{L_\text{I}}{4 G_N} \log \left[\frac{\sigma^2+y^2+2 \sigma \,y \sin \psi_{\rm I}}{\epsilon \, y \cos \psi_{\rm I}}\right]+ \frac{L_\text{II}}{4 G_N} \log \left[\frac{\sigma^2+y^2+2 \sigma \,y \sin \psi_{\rm II}}{\epsilon \, y \cos \psi_{\rm II}}\right],
\end{align}
where we have considered $\sigma_1=\sigma_2=\sigma$ for simplicity. Now, to obtain the holographic entanglement entropy, we extremize the above with respect to $y$ to obtain
\begin{align}\label{ext_zeroT_semi}
	L_\text{II}\, y (y-\sigma ) (y+\sigma )& \left[\left(y^2+\sigma ^2\right) \left(\cos \psi _\text{I}\, \sec \psi _\text{II}+1\right)+2 \sigma \, y \sin (\psi _\text{I}+\psi _\text{II}) \sec \psi _\text{II}\right]\notag\\
	& -4 G_N \Phi _r \left(y^2+\sigma ^2+2 \sigma \, y \sin \psi _\text{I}\right) \left(y^2+\sigma ^2+2 \sigma \, y \sin \psi _\text{II}\right)=0~.
\end{align}
The entanglement entropy for the semi-infinite subsystem in question may be obtained by substituting the physical solution for $y$ in \cref{EE}. Finally, in the large tension limit described in \cref{tensionlimit}, this matches exactly with the corresponding entanglement entropy computed in the effective $2d$ theory on utilization of the Brown-Henneaux formula $c_\text{I,II}=\frac{3 {L}_\text{I,II}}{2 G_N}$ \cite{Brown:1986nw}.


\subsection{Finite subsystem}\label{sec:double-crossing}
In this subsection, we obtain the entanglement entropy for a finite sized subsystem $A \equiv [\sigma_1,\sigma_2]_\text{I} \cup [\sigma_1,\sigma_2]_\text{II}$ located in the baths CFT$_2 ^{\text{I}}$ and CFT$_2 ^{\text{II}}$. Here, we observe three non-trivial phases for the generalized entanglement entropy depending upon the sizes of the subsystem $A$ as depicted in \cref{fig:zeroTdouble,fig:zeroTsingle,fig:zeroTdome}. In this context, we first utilize the effective $2d$ prescription to compute the generalized entanglement entropy for the corresponding subsystem in these scenarios. Subsequently, we provide a doubly holographic characterization of the entanglement entropy for the three cases using the RT prescription which substantiates the corresponding field theory results. 

\subsubsection{Phase - I} \label{sec:finite-size-I}
\subsection*{Effective $2d$ description}

\begin{figure}[ht]
	\centering
	\includegraphics[scale=0.8]{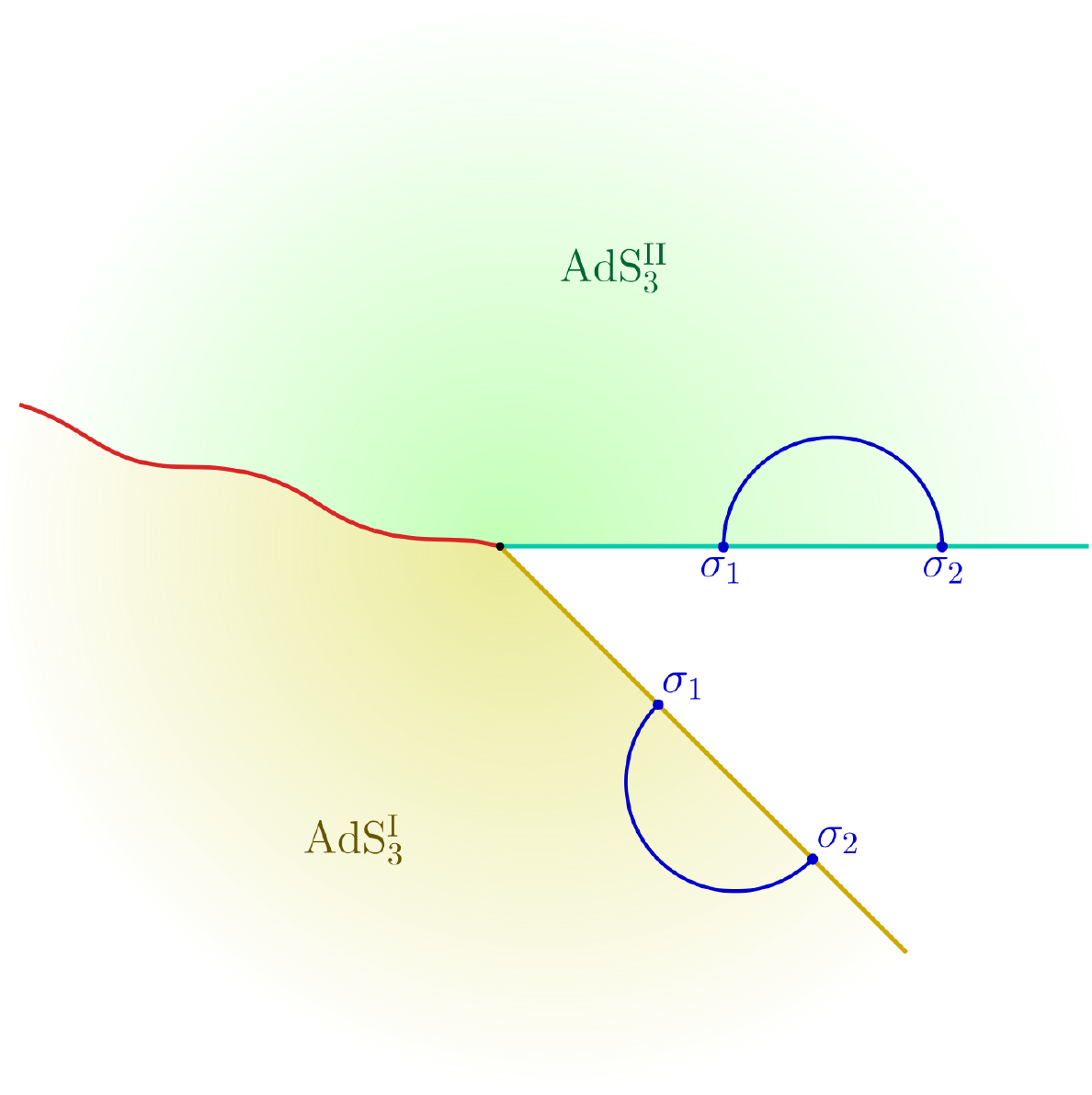}
	\caption{Schematics of the phase-I for the bulk geodesic homologous to the subsystem $A$ described by a finite interval $[\sigma_1,\sigma_2]_\text{I,II}$ in dual CFT$^\text{I,II}$.}
	\label{fig:zeroTdome}
\end{figure}

We begin with the computation of the generalized entanglement entropy for the phase where the intervals $[\sigma_1,\sigma_2]_\text{I}$ and $[\sigma_1,\sigma_2]_\text{II}$ are small such that no island region is observed as depicted in \cref{fig:zeroTdome}. For this configuration the area term in generalized entropy vanishes and the expression for the entanglement entropy may trivially be obtained to be
\begin{equation}\label{sgenzeroTdome}
	\begin{aligned}
		S_A	=&\frac{(c_\text{I}+c_\text{II})}{3} \log \left[\frac{ \sigma_2-\sigma_1}{\epsilon}\right] \,.
	\end{aligned}
\end{equation}

\subsection*{Doubly holographic description}
From the doubly holographic perspective, it may be observed that the RT surfaces for the intervals $[\sigma_1,\sigma_2]_\text{I,II}$ in CFT$_2^\text{I,II}$ are described by the usual dome-shaped geodesics each in the dual bulk AdS$_3^\text{I,II}$ geometries as depicted in \cref{fig:zeroTdome}. The entanglement entropy for this configuration may then be obtained to be
\begin{align}
	S_{\rm dome} = \frac{(L_\text{I}+L_\text{II})}{4G_N} \log \left[\frac{ \sigma_2-\sigma_1}{\epsilon}\right] \,,
\end{align}
which matches identically with the corresponding expression obtained in the effective $2d$ description in \cref{sgenzeroTdome} through the utilization of the Brown-Henneaux formula.

\subsubsection{Phase - II}\label{sec:finite-size-II}
\subsection*{Effective $2d$ description}
We now discuss the next phase where the sizes of the intervals in the CFT$_\text{I,II}$ are increased such that we now observe an island region on the JT brane described by $[-a_2,-a_1]_\text{I,II}$. Note that this configuration corresponds to the conventional origin of the island as discussed in \cref{sec:S-gen}. The effective terms of the generalized entanglement entropy given in \cref{conventional-Sgen} for this case may be obtained through the four-point twist correlators in CFT$_2^\text{I,II}$ which factorize in the large-$c$ limit in the following way
\begin{equation}\label{sgenzeroTfinite-factorization}
	\begin{aligned}
		\left<\mathcal{T}_n(\sigma_1) \bar{\mathcal{T}}_n(\sigma_2)\mathcal{T}_n(-a_2) \bar{\mathcal{T}}_n(-a_1)\right>_{\text{CFT}^{{k}}_{2}} = \left<\mathcal{T}_n(\sigma_1) \bar{\mathcal{T}}_n(-a_1)\right>_{\text{CFT}^{{k}}_{2}} \left<\bar{\mathcal{T}}_n(\sigma_2) \mathcal{T}_n(-a_2)\right>_{\text{CFT}^{{k}}_{2}} \,.
	\end{aligned}
\end{equation}
Subsequently, we may express the generalized entropy in the large-tension limit as follows,
\begin{equation}
	\begin{aligned}
		S_{\text{gen}}=&\frac{\Phi_r}{a_1}+\frac{c_\text{I}}{6}\log \left[\left(\frac{L_\text{I}+L_\text{II}}{L_\text{I}\,\delta
		}\right)\frac{\left(\sigma _1+a_1\right)^2}{a_1}\right]+\frac{c_\text{II}}{6}\log \left[\left(\frac{L_\text{I}+L_\text{II}}{L_\text{II}\,\delta
		}\right)\frac{\left(\sigma_1+a_1\right)^2}{a_1}\right]\\
		&+\frac{\Phi_r}{a_2}+\frac{c_\text{I}}{6}\log \left[\left(\frac{L_\text{I}+L_\text{II}}{L_\text{I}\,\delta
		}\right)\frac{\left(\sigma _2+a_2\right)^2}{a_2}\right]+\frac{c_\text{II}}{6}\log \left[\left(\frac{L_\text{I}+L_\text{II}}{L_\text{II}\,\delta
		}\right)\frac{\left(\sigma_2+a_2\right)^2}{a_2}\right]\,.
	\end{aligned}
\end{equation}
Similar to \cref{sec:single-crossing}, the above may be extremized over the positions of the QES at $a_1$ and $a_2$ to obtain expressions similar to \cref{a-extr-inf-T-0} with $\sigma$ replaced by $\sigma_1$ and $\sigma_2$ respectively. Finally, the entanglement entropy for this configuration may be obtained by substituting these extremal values of the island locations in the above generalized entropy.

\begin{figure}[ht]
	\centering
	\includegraphics[scale=0.75]{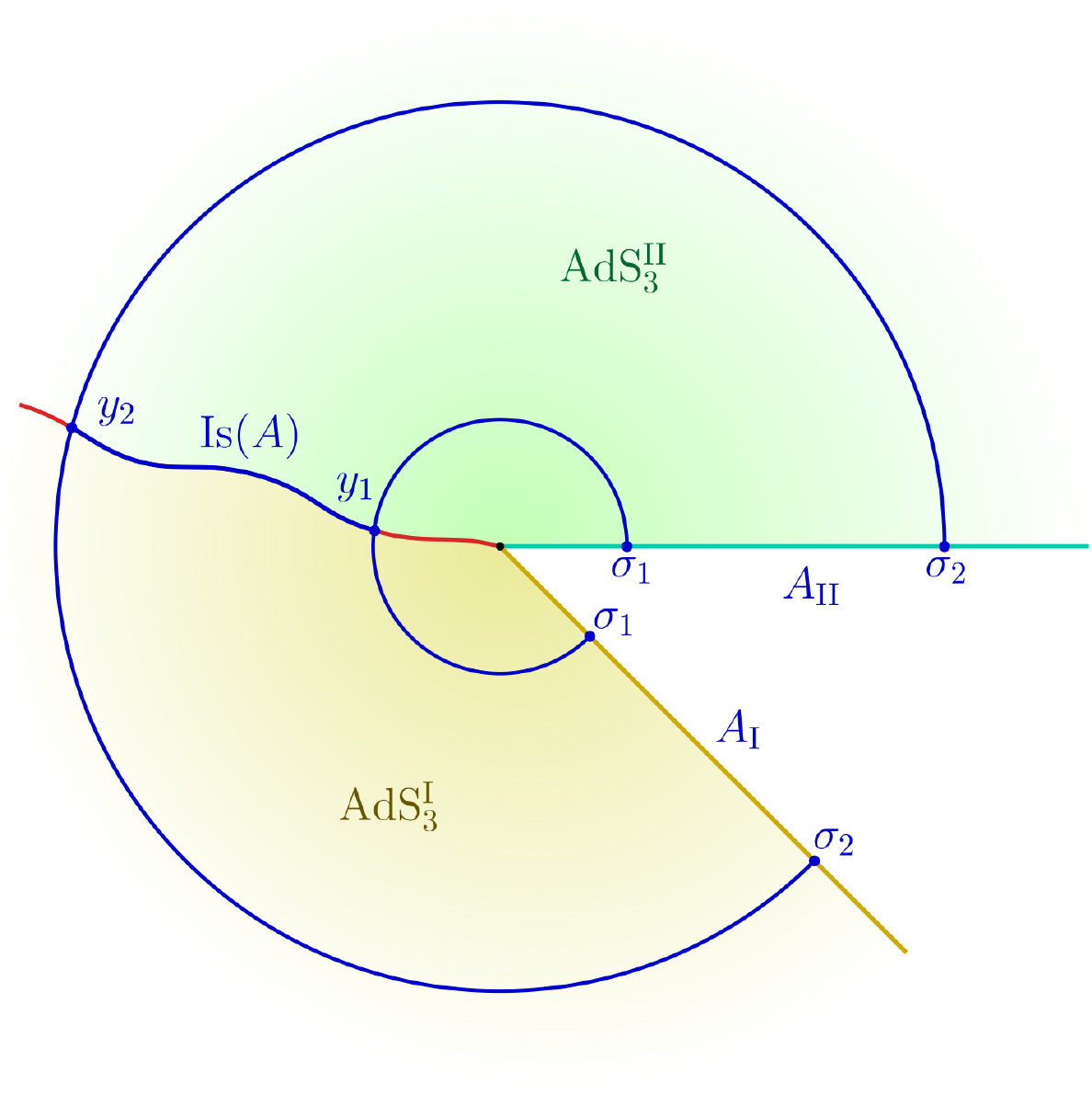}
	\caption{Schematics of the phase-II for the bulk geodesic homologous to the subsystem $A$ described by a finite interval $[\sigma_1,\sigma_2]_\text{I,II}$ in dual CFT$^\text{I,II}$. Here Is$(A)$ denotes the island region in the effective $2d$ description.}
	\label{fig:zeroTsingle}
\end{figure}

\subsection*{Doubly holographic description}
For this phase, owing to the conventional island, the RT surface homologous to subsystem $A$ is composed of two single-crossing geodesics of the type discussed in \cref{sec:single-crossing} and as depicted in \cref{fig:zeroTsingle}. Consequently, the candidate entanglement entropy for the present configuration may be expressed as
\begin{equation}\label{EEfinitesize}
	\begin{aligned}
		S_{\text{bulk}}=S_{\text{single}} (\sigma_1,y_1) + S_{\text{single}} (\sigma_2,y_2) \,.
	\end{aligned}
\end{equation}
As earlier, to obtain the locations of the common points $y_i$s on the EOW brane, we extremize the above with respect to $y_1$ and $y_2$ to obtain equations analogous to \cref{ext_zeroT_semi} whose physical solutions will lead to the holographic entanglement entropy. Subsequently in the large tension limit described by \cref{tensionlimit}, this agrees with the corresponding result in the effective $2d$ description.

\subsubsection{Phase - III}\label{sec:finite-size-III}

\subsection*{Effective $2d$ description}
\begin{figure}[ht]
	\centering
	\includegraphics[scale=0.8]{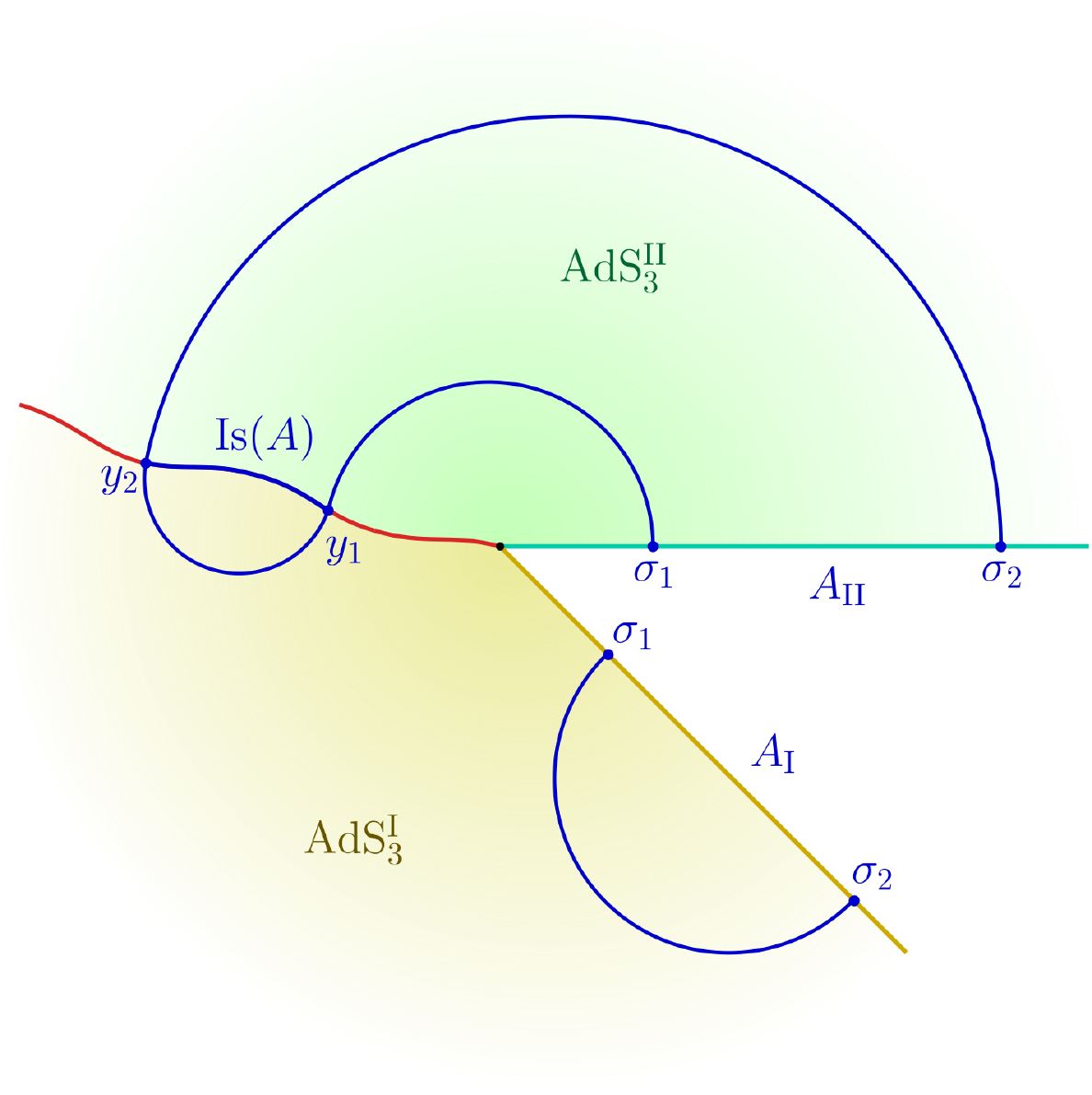}
	\caption{Schematics of the phase-III for the bulk geodesic homologous to the subsystem $A$ described by the finite interval $[\sigma_1,\sigma_2]_\text{I,II}$ in dual CFT$^\text{I,II}$. Here Is$(A)$ denotes the \textit{induced} island region in the effective $2d$ description.}
	\label{fig:zeroTdouble}
\end{figure}

We now proceed to the final phase depicted in \cref{fig:zeroTdouble} which involves the novel induced islands discussed in \cref{sec:S-gen}. The location of this island region is described by $[a_1,a_2]_{\text{I,II}}$ on the JT brane for the given subsystem. For this phase, we utilize the generalized entanglement entropy formula in \cref{novel-Sgen} for the corresponding subsystem $A$. The presence of induced island results in the factorization of the four-point twist correlators in the effective terms of the generalized entanglement entropy in the following way
\begin{equation}\label{sgenfinite-factorization}
	\begin{aligned}
		\left<\mathcal{T}_n(\sigma_1) \bar{\mathcal{T}}_n(\sigma_2)\mathcal{T}_n(-a_2) \bar{\mathcal{T}}_n(-a_1)\right>_{\text{CFT}^{\text{I}}_{2}} &= \left<\mathcal{T}_n(\sigma_1) \bar{\mathcal{T}}_n(\sigma_2)\right>_{\text{CFT}^{\text{I}}_{2}} \left<\mathcal{T}_n(-a_1) \bar{\mathcal{T}}_n(-a_2)\right>_{\text{CFT}^{\text{I}}_{2}} \\
		\left<\mathcal{T}_n(\sigma_1) \bar{\mathcal{T}}_n(\sigma_2)\mathcal{T}_n(-a_2) \bar{\mathcal{T}}_n(-a_1)\right>_{\text{CFT}^{\text{II}}_{2}} &= \left<\mathcal{T}_n(\sigma_1) \bar{\mathcal{T}}_n(-a_1)\right>_{\text{CFT}^{\text{II}}_{2}} \left<\mathcal{T}_n(\sigma_2) \bar{\mathcal{T}}_n(-a_2)\right>_{\text{CFT}^{\text{II}}_{2}} \,.
	\end{aligned}
\end{equation}
On utilization of the area term given in \cref{area-term} and the position dependent cut-off $\epsilon(y)$ on the JT brane, the generalized entanglement entropy in this case may then be expressed as
\begin{align}\label{finitesgen}
	S_\text{gen}=\frac{\Phi_r}{a_1}+\frac{\Phi_r}{a_2}+\frac{c_{\text{I}}}{3}\log\left[\frac{1}{\cos\psi_{\text{I}}}\right]+\frac{c_{\text{II}}}{3}\log\left[\frac{1}{\cos\psi_{\text{II}}}\right]+\frac{c_\text{I}}{6}\log
	\left[\frac{ \left(\sigma_2-\sigma_1\right)}{\epsilon }\right]\\ \nonumber
	+\frac{c_\text{I}}{6} \log \left[\frac{ \left(a_1-a_2\right)^2}{a_1\, a_2}\right] +\frac{c_\text{II}}{6} \log \left[\frac{ \left(\sigma_1+a_1\right)^2}{\epsilon\, a_1}\right]+\frac{c_\text{II}}{6}\log
	\left[\frac{ \left(\sigma_2+a_2\right)^2}{\epsilon\, a_2}\right]
	\,.
\end{align}

We now introduce a parameter $\Theta=\frac{\sigma_2}{\sigma_1}$ which is motivated from the analysis\footnote{Note that the authors \cite{Anous:2022wqh} only described the bulk computation of the entanglement entropy for finite interval located in the CFT$_2 ^\text{II}$.} described in \cite{Anous:2022wqh}. Moreover, one can also establish a similar relation between the QES $a_1$ and $a_2$ located on the JT brane as $a_2 = {\kappa} \, \Theta \, a_1$ where $\kappa$ is now one of the parameters whose extremal value will minimize the entanglement entropy. In the case of non-perturbed EOW brane where we obtain the usual ICFT$_2$ setup, it was shown in \cite{Anous:2022wqh} that the current phase is only possible above a certain value of the parameter $\Theta$ depending upon the configuration of the EOW brane. Thus in our computations we assume $\Theta$ to be large. Finally, we introduce $\Theta$ and $\kappa$ in \cref{finitesgen}, and extremize over the parameters $a_1$ and $\kappa$ in the large tension limit to obtain the following relations,
\begin{equation}
	\begin{aligned}
		\partial_{\kappa} S_\text{gen}=0\quad \Rightarrow \quad \quad	&a_1 \left(c_\text{I}+c_\text{II}\right) \kappa+\left(c_\text{I}-c_\text{II}\right) \sigma _1=0\\ 
		\partial_{a_1} S_\text{gen}=0\quad \Rightarrow \quad \quad& a_1 \sigma _1 \left(c_\text{II} \sigma _1+3 (\kappa+1) \Phi_r\right) - a_1^3 \kappa\, c_\text{II} +3 a_1^2 \kappa\, \Phi_r+3 \sigma _1^2 \Phi_r=0\,.
	\end{aligned}
\end{equation}
Solving the above, the extremal values of the QES are obtained to be
\begin{equation}
	\begin{aligned}
		\kappa ^*=&\frac{\left(c_\text{II}-c_\text{I}\right) \, \sigma_1}{\left(c_\text{I}+c_\text{II}\right)\, a_1^*} \, \\
		a_1^*=&\frac{\left(c_\text{I}+c_\text{II}\right) \sigma _1+6 \Phi_r+\sqrt{\left(\left(c_\text{I}+c_\text{II}\right) \sigma _1+6 \Phi_r\right)^2+24 \left(c_\text{II}-c_\text{I}\right) \sigma _1 \Phi_r}}{2 \left(c_\text{II}-c_\text{I}\right)}
	\end{aligned}
\end{equation}
where we have only considered the physical solutions of the island surfaces. The fine grained entropy for the subsystem $A$ may now be obtained by substituting the above extremal value in the generalized entropy in \cref{finitesgen}.

\subsection*{Doubly holographic description}
In this subsection we obtain the length of the RT surfaces supported by the finite-sized subsystem $[\sigma_1, \sigma_2]_{\rm I,II}$ located in the dual $\text{CFT}_2^{\rm I,II}$ as depicted in \cref{fig:zeroTdouble}. The interval in CFT$_2^\text{I}$ supports the usual boundary anchored dome-shaped geodesic. However, for the interval in CFT$_2^\text{II}$, the extremal curve is composed of three circular segments forming a double-crossing RT saddle as discussed in \cref{sec:S-gen}. This double-crossing geodesic intersects the EOW brane at $y_1$ and $y_2$ which form the boundary of the island region in the effective $2d$ description.

Utilizing \cref{geoleninpoin}, the length of the double-crossing geodesic may be obtained as\footnote{Note that the transverse fluctuation of the EOW brane requires the brane angles $\psi_{\rm I,II} (y_i)$ to be position dependent. }
\begin{align}
	d=&L_\text{I} \cosh ^{-1}\left[\frac{\left(y_2-y_1\right){}^2 \sin \psi_\text{I}(y_1)\sin \psi_\text{I}(y_2)+\left(y_1^2+y_2^2\right) \cos \psi_\text{I}(y_1)\cos \psi_\text{I}(y_2)}{2 y_1 y_2 \cos \psi_\text{I}(y_1)\cos \psi_\text{I}(y_2)}\right]\notag \\ &+L_\text{II} \log \left[\frac{\left(\sigma _1+y_1 \sin \psi_\text{II}(y_1)\right){}^2+\left(y_1 \cos \psi_\text{II}(y_1)\right){}^2}{y_2 \cos \psi_\text{II}(y_1)}\right]\notag \\&+L_\text{II} \log \left[\frac{\left(\sigma _2+y_2 \sin \psi_\text{II}(y_2)\right){}^2+\left(y_2 \cos \psi_\text{II}(y_2)\right){}^2}{y_2 \cos \psi_\text{II}(y_2)}\right]\,.
\end{align}
We may now introduce the variables $\Theta=\frac{\sigma_2}{\sigma_1}$ and $y_2 = \hat\kappa \, \Theta \, y_1$, similar to the effective $2d$ perspective considered earlier. As advocated in \cite{Anous:2022wqh} in the context of AdS$_3$/ICFT$_2$, such double-crossing geodesics are only permissible for large $\Theta$s. Consequently, in the large $\Theta$ limit of the above length, we obtain
\begin{equation} \label{nunu}
	\begin{aligned}
		d \approx L_\text{I} \log \left[\frac{\Theta \, \hat\kappa}{\cos \psi _{\rm I}(y_1) \cos \psi _{\rm I}(y_2)} \right]&+L_\text{II} \log \left[\frac{\sigma_1^2+2 y_1 \sigma_1 \sin \psi_{\rm II}(y_1)+y_1^2}{y_1 \cos \psi _{\rm II}(y_1)}\right]\\
		&+L_\text{II} \log \left[\frac{\Theta (\sigma_1^2+2 \hat\kappa y_1 \sigma_1 \sin \psi_{\rm II}(y_2)+\hat\kappa^2 y_1^2)}{\hat\kappa \, y_1 \cos \psi _{\rm II}(y_2)}\right] \,.
	\end{aligned}
\end{equation}
Next we implement the position dependence of the brane angles $\psi_{\rm I,II} (y_i)$ explicitly in the following way
\begin{equation}
	\begin{aligned}
		\psi_{\rm I}(y_i) &\to \sin ^{-1}\left[\tanh \left(\frac{\rho^0_{\rm I}-\tilde{\rho}_{\rm I}\left(y_i\right)}{L_{\rm I}}\right)\right] \,,\\
		\psi_{\rm II}(y_i) &\to \sin ^{-1}\left[\tanh \left(\frac{\rho^0_{\rm II}+\tilde{\rho}_{\rm II}\left(y_i\right)}{L_{\rm II}}\right)\right]~.
	\end{aligned}
\end{equation}
Expanding \cref{nunu} upto the leading order in $\tilde \rho_\text{I,II}$ and identifying the dilaton as in \cref{phi_y}, we may obtain the corresponding contribution from the double-crossing geodesic as follows
\begin{align}\label{finitesbulk}
	&S_\text{double} (y_1,y_2)= \frac{\Phi_r}{y_2}+\frac{\Phi_r}{y_1}+\frac{L_\text{I}}{4 G_N} \log \left[\frac{y_2}{y_1 } \sec ^2\left(\psi _{\rm I}\right)\right]\notag\\
	&+\frac{L_\text{II}}{4 G_N}\log \left[\frac{ \left(\sigma _2^2+2 y_2 \, \sigma _2 \sin \psi _{\rm II}+y_2^2\right) \left(\sigma _1^2+2 y_1 \, \sigma _1 \sin \psi _{\rm II}+y_1^2\right)}{y_1 \, y_2 \cos ^2\psi _{\rm II}}\right]~,
\end{align}
where we have restored the original variables $\sigma_2$ and $y_2$.

Finally the candidate entanglement entropy for finite sized subsystem under consideration may be obtained by including the contribution from the dome-shaped geodesic as follows
\begin{align}\label{finitesbulk3}
	S_\text{bulk}&= S_\text{double} (y_1,y_2)+\frac{L_\text{I}}{4 G_N}\log
	\left[\frac{ \left(\sigma_2-\sigma_1\right)}{\epsilon\,\, }\right] \,.
\end{align}
The above may be extremized over the undetermined parameters $y_1$ and $y_2$ to obtain
\begin{align}
	2 y_1 \, \sigma _1 \sin \psi _{\rm II} \left(4 G_N \Phi_r  \sin \psi _{\rm I}+L_{\rm I} y_1\right)	&+\sigma _1^2 \left(4 G_N \Phi_r  \sin \left(\psi _{\rm I}\right)+\left(L_{\rm I}+L_{\rm II}\right) y_1\right)\notag\\
	&\quad-y_1^2 \left(\left(L_{\rm II}-L_{\rm I}\right) y_1-4 G_N \Phi_r  \sin \psi _{\rm I}\right)=0~,\label{ext1}\\
	2 y_2 \, \sigma _2 \sin \psi _{\rm II} \left(4 G_N \Phi_r  \sin \psi _{\rm I}-L_{\rm I} y_2\right)&+\sigma _2^2 \left(4 G_N \Phi_r  \sin \psi _{\rm I}-\left(L_{\rm I}-L_{\rm II}\right) y_2\right)\notag \\
	&\quad -y_2^2 \left(\left(L_{\rm I}+L_{\rm II}\right) y_2-4 G_N \Phi_r  \sin \psi _{\rm I}\right)=0~.\label{ext2}
\end{align}
Solving the above equations for $y_1$ and $y_2$ and substituting the extremal values in \cref{finitesbulk3} will finally result in the holographic entanglement entropy for the given finite subsystem. Once again, in the large tension limit described in \cref{tensionlimit}, we observe that the corresponding entanglement entropy obtained through the effective $2d$ description is reproduced.

\subsubsection{Page curve}
\begin{figure}[H]
	\centering
	\includegraphics[scale=0.565]{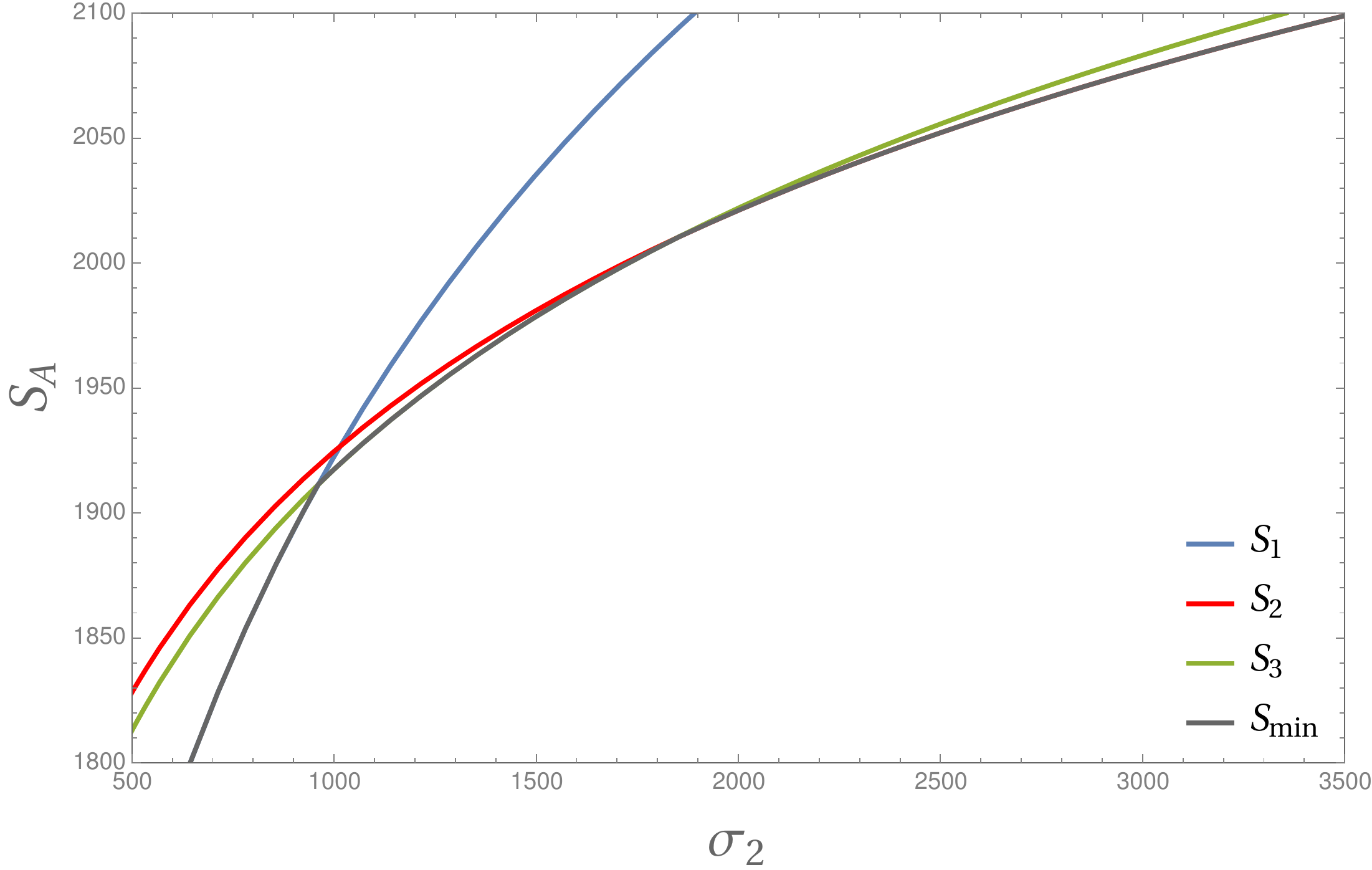}
	\caption{Variation of the entanglement entropy with the subsystem size ($\sigma_2$) for the finite sized subsystem $A$ where $c_\text{I}=35,c_\text{II}=800, \Phi_r=6.958, \delta=0.1, \sigma_1=0.4$. Here $S_1,\,S_2,\,S_3$ correspond to the entropy in phase-I, phase-II and phase-III respectively. The entanglement entropy is given by the minimum $S_\text{min}$ of these possible candidates for a given subsystem size.}
	\label{fig:transition}
\end{figure}

We now plot the entanglement entropy for the finite subsystem $A$ under consideration in the dual CFT$_2 ^{\text{I,II}}$s with respect to the subsystem size in \cref{fig:transition}. For the given value of parameters, we observe transitions between the three phases discussed above as the subsystem size is increased. Initially when the subsystem is small in size, \hyperref[sec:finite-size-I]{phase-I} has the minimum entanglement entropy and is dominant. As we increase the subsystem size, subsequently \hyperref[sec:finite-size-III]{phase-III} starts dominating as crossing over to the region with smaller AdS radius AdS$_3^\text{I}$, is more economical for the geodesic. Finally if the subsystem size is further increased, this advantage of double-crossing vanishes as the length of dome-shaped RT surface supported by the interval in CFT$_2^\text{I}$ keeps increasing. And ultimately, \hyperref[sec:finite-size-II]{phase-II} becomes dominant.

\section{Islands outside eternal JT black holes }\label{sec:FiniteT}

In this section, we consider semi-infinite subsystems in thermal CFT$_2$ baths coupled to an eternal JT black hole. The thermofield double (TFD) state in this case may be constructed through the Euclidean path integral on half of an infinite cylinder \cite{Anous:2022wqh}. The corresponding cylinder geometry may be obtained by applying a series of transformations on the planar ICFT$_2$ setup described by $\zeta = x + i t_E$. We begin by mapping the flat interface to a circle of length $\ell$ through the following SL(2,$\mathbb{R}$) transformation 
\begin{align}
	p = \frac{4 \ell^2 }{2 \ell - \zeta} - \ell \, ,
\end{align}
where $p=\tilde{x}+i\tilde{t}_E$. The corresponding bulk transformations may be obtained through the Ba\~nados formalism \cite{Banados:1998gg, Roberts:2012aq, Shimaji:2018czt} as follows
\begin{equation}\label{Barnados}
	\begin{aligned}
		\tilde{x}=\frac{x -\frac{x^2+z^2 -t^2 }{2\ell}}{1-\frac{x}{\ell}+\frac{x^2 +z^2 -t^2 }{4\ell^2}}+\ell\,,\qquad \tilde{z}= \frac{z}{1-\frac{x}{\ell} +\frac{x^2 +z^2 -t^2 }{4\ell^2}}\,, \qquad \tilde{t}= \frac{t}{1-\frac{x}{\ell} +\frac{x^2 +z^2 -t^2 }{4\ell^2}}\,.
	\end{aligned}
\end{equation}
We further obtain the cylinder geometry via the usual exponential map given by
\begin{equation} \label{cyl-map}
	p=\ell e^{\frac{2 \pi}{\beta} q}~,
\end{equation}
where the coordinate $q=u+iv_E$ describes the cylinder with circumference $\beta$. The interface is now mapped to a circle $\mathfrak{R}\mathfrak{e} (q) = 0$ with the two CFT$_2$s mapped on either side. The dual bulk theory for the TFD state on this cylinder is then described by an eternal black string spanning two AdS$_3$ geometries separated by a thin AdS$_2$ brane. The horizon of the black string crosses the brane and induces a horizon on it. A similar partial dimensional reduction as described in \cref{sec:JT-EOW} may now be performed for this $3d$ bulk to obtain an effective $2d$ description comprising of two thermal CFT$_2$ baths coupled to an eternal JT black hole.

In the cylinder coordinates, the metric and the dilaton profile for the eternal JT black hole in AdS$_2$ are given as follows \cite{Almheiri:2019hni, Almheiri:2019qdq, Almheiri:2019yqk} 
\begin{align}\label{phi_T}
	&\textrm{d}s_{\text{grav}}^2=\frac{4\pi^2}{\beta^2}\frac{\text{d}q\,\text{d}\bar{q}}{\sinh^2\left(\frac{\pi(q+\bar{q})}{\beta}\right)}\quad,
	&\Phi=\Phi_0-\frac{2 \pi  \Phi_r }{\beta }\coth \left(\frac{\pi(q+\bar{q})}{\beta}\right)\,,
\end{align}
where $\Phi_0$ is the topological contribution to the dilaton given in \cref{area-term}. On the other hand, the metric for the CFT$_2$ baths may be expressed as
\begin{align}\label{bath-metric}
	&\textrm{d}s_{\text{bath}}^2=\frac{1}{\epsilon^2}\text{d}q\,\text{d}\bar{q}\,.
\end{align}

However, in the following we will employ the planar coordinates $p$ in which the field theory remains in the ground state and the corresponding stress tensor vanishes. The corresponding metrics and the dilaton are given as follows
\begin{flalign} \label{planar-dilaton}
	&\textrm{d}s_{\text{grav}}^2=\frac{4}{\left(1-|p|^2\right)^2}\text{d}p\,\text{d}\bar{p}\quad,\quad \textrm{d}s_{\text{bath}}^2=\frac{\beta^2}{4\pi^2\epsilon^2}\frac{\text{d}p\,\text{d}\bar{p}}{|p|^2}\,,\notag\\
	&\qquad\qquad\qquad\Phi=\Phi_0+\frac{2 \pi  \Phi_r }{\beta }\frac{1+|p|^2}{1-|p|^2} \,.
\end{flalign}

We will now obtain the fine-grained entanglement entropy for a semi-infinite subsystem in bath CFT$_2^{\text{I,II}}$s coupled to an eternal JT black hole. For this case, we observe two phases for the entanglement entropy as depicted in \cref{fig:semi_infiniteT-HM,fig:semi_infiniteT-island}. Specifically, for the first phase, we do not observe islands and obtain a steadily rising entanglement entropy as the black hole evolves. For the second phase, QES are observed outside the eternal JT black holes, indicating the presence of islands which saturates the entanglement entropy.

\subsection{Phase - I}\label{sec:FT-I}
\subsection*{Effective $2d$ description}
\begin{figure}[ht]
	\centering
	\includegraphics[scale=0.8]{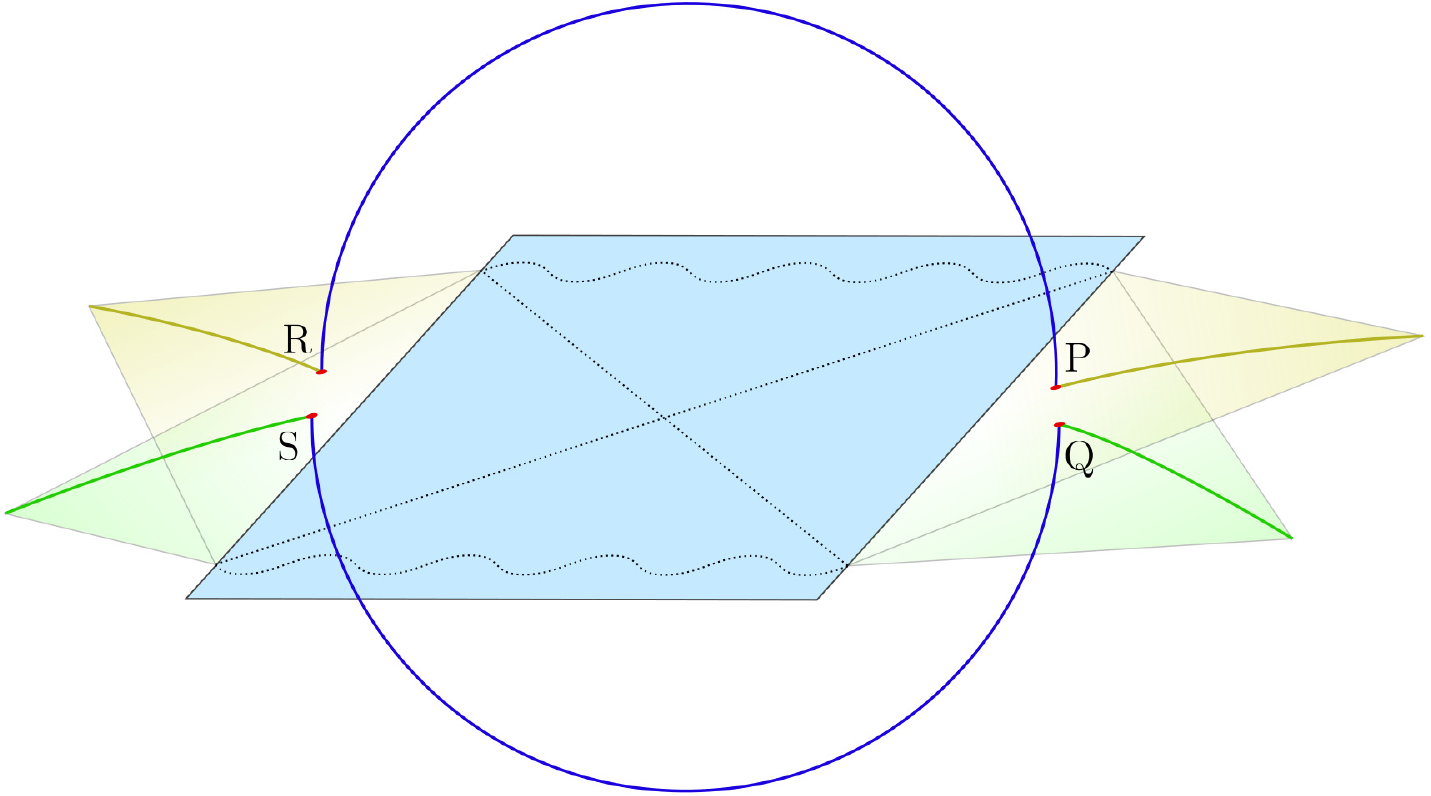}
	\caption{Schematics for the semi-infinite subsystem $A$ in phase-I where the extremal curve is composed of two Hartman-Maldacena surfaces.}
	\label{fig:semi_infiniteT-HM}
\end{figure}

Now we describe the computation of the generalized entanglement entropy in the first phase for the subsystem composed of semi-infinite intervals  $[P,\infty]_\text{I}\cup[R,\infty]_\text{I}$ considered in CFT$_2^{\text{I}}$ bath and $[Q,\infty]_\text{II}\cup[S,\infty]_\text{II}$ in CFT$_2^{\text{II}}$ bath as depicted in the \cref{fig:semi_infiniteT-HM}. Here the points $P$, $Q$ have coordinates as $(u_0,v)_\text{I,II}$ in the cylinder coordinates and the points $R$, $S$ are their corresponding TFD copies with coordinates $(u_0,-v+i\frac{\beta}{2})_\text{I,II}$. Note that in this case, the area term in the generalized entanglement entropy formula is vanishing as no island region is observed for this phase. The generalized entropy then involves only the effective term described by two two-point twist correlators and may be obtained to be 
\begin{equation}
	\begin{aligned}
		S_{A}=	\frac{ (c_\text{I}+c_\text{II})}{3}  \log \left[\frac{\beta  \cosh \left(\frac{2 \pi  v}{\beta }\right)}{\pi \, \epsilon
		}\right]\,.
	\end{aligned}
\end{equation}

\subsection*{Doubly holographic description}
The double holographic description for this phase corresponds to the RT surfaces being composed of two Hartman-Maldacena (HM) surfaces stretched between the endpoints of the semi-infinite intervals on the asymptotic boundaries as depicted in \cref{fig:semi_infiniteT-HM}. The endpoints of the intervals are specified in the planer coordinates as $\tilde{x}_1(q_1^{\rm I})$ and $\tilde{x}_1(q_1^{\rm II})$ which may be obtained via the conformal maps \cref{Barnados,cyl-map}. Consequently, in this phase the entanglement entropy corresponding to the HM surfaces may be obtained as
\begin{align}\label{Sbulkhm}
	S_{\rm bulk} &= \frac{L_\text{I}}{2G_N} \log\left[ \frac{2 \tilde{x}_1(q_1^{\rm I})}{\tilde{\epsilon}} \right] + \frac{L_\text{II}}{2G_N} \log\left[ \frac{2 \tilde{x}_1(q_1^{\rm II})}{\tilde{\epsilon}} \right]\notag\\
	&= \frac{ (L_\text{I}+L_\text{II})}{2 G_N}  \log \left[\frac{\beta  \cosh \left(\frac{2 \pi  v}{\beta }\right)}{\pi  \epsilon
	}\right]\,.
\end{align}
Note that the UV cut-offs between the two coordinates are related by $\tilde{\epsilon} (u,v) = \epsilon \,\frac{2 \pi \ell}{\beta} e^\frac{2  \pi u}{\beta} $. The above expression matches identically with the result obtained in the effective $2d$ description.

\subsection{Phase - II}\label{sec:FT-II}
\subsection*{Effective $2d$ description}
Now we describe the second phase for the generalized entropy of semi-infinite subsystems $[P,\infty]_\text{I}\cup[R,\infty]_\text{I}$ in CFT$_2 ^{\text{I}}$ bath and $[Q,\infty]_\text{II}\cup[S,\infty]_\text{II}$ in CFT$_2 ^{\text{II}}$ bath. Note that, as earlier, the points $P$, $Q$ are located at $(u_0,v)_\text{I,II}$ in the cylinder coordinates and the points $R$, $S$ are their corresponding TFD copies. This phase involves a conventional island region bounded by the QES $M \equiv (-a,v^{}_a)_\text{I,II}$ and $N \equiv (-a,-v^{}_a+i\frac{\beta}{2})_\text{I,II}$ on the JT brane leading to area terms in the generalized entanglement entropy. The effective terms in \cref{conventional-Sgen} now involve four two-point twist correlators. The generalized entanglement entropy may then be obtained as 
\begin{equation}\label{Siland}
	\begin{aligned}
		S_\text{gen}=&\frac{4 \pi  \Phi_r }{\beta }\coth \left(\frac{2 \pi a}{\beta }\right)+\frac{c_{\text{I}}}{3}\log\left[\frac{1}{\cos\psi_{\text{I}}}\right]+\frac{c_{\text{II}}}{3}\log\left[\frac{1}{\cos\psi_{\text{II}}}\right]\\
		&+\frac{(c_\text{I}+c_\text{II})}{3} \log \left[\frac{\left( e^{\frac{2 \pi (-a-v^{}_a)}{\beta} } - e^{\frac{2 \pi (u_0-v)}{\beta} }\right) \left( e^{\frac{2 \pi (-a+v^{}_a)}{\beta} } - e^{\frac{2 \pi (u_0+v)}{\beta} }\right)}{\frac{\pi\, \epsilon}{\beta} e^{\frac{2 \pi u}{\beta}} \left( 1 - e^{\frac{- 4 \pi a}{\beta}}\right)}\right]\,.
	\end{aligned}
\end{equation}
We first extremize the above over the time $v^{}_a$ of the QES to obtain the extremal value as
\begin{equation}\label{extcondisland}
	\begin{aligned}
	\partial_{v^{}_a} S_\text{gen}=0\quad &\Rightarrow  \quad \quad v^{*}_a = v\,.
	\end{aligned}
\end{equation}
Subsequently, the extremization of the generalized entropy is performed over the location $a$ of the QES to obtain the following equation
\begin{equation}\label{extcondisland2}
	\begin{aligned}
		\partial_a S_\text{gen}=0\quad \Rightarrow \quad		&\frac{\sinh \left(\frac{\pi  (a-u)}{\beta }\right)}{\sinh \left(\frac{\pi  (a+u)}{\beta }\right)}=\frac{12 \pi  \Phi_r }{\beta  \left(c_\text{I}+c_\text{II}\right)}\text{csch}\left(\frac{2 \pi  a}{\beta }\right) \,.
	\end{aligned}
\end{equation}
where we have implemented $v^{*}_a = v$. The fine grained entanglement entropy for this configuration may be obtained by solving the above for the extremal value $a^*$ and substituting it in \cref{Siland}.

\begin{figure}[ht]
	\centering
	\includegraphics[scale=0.8]{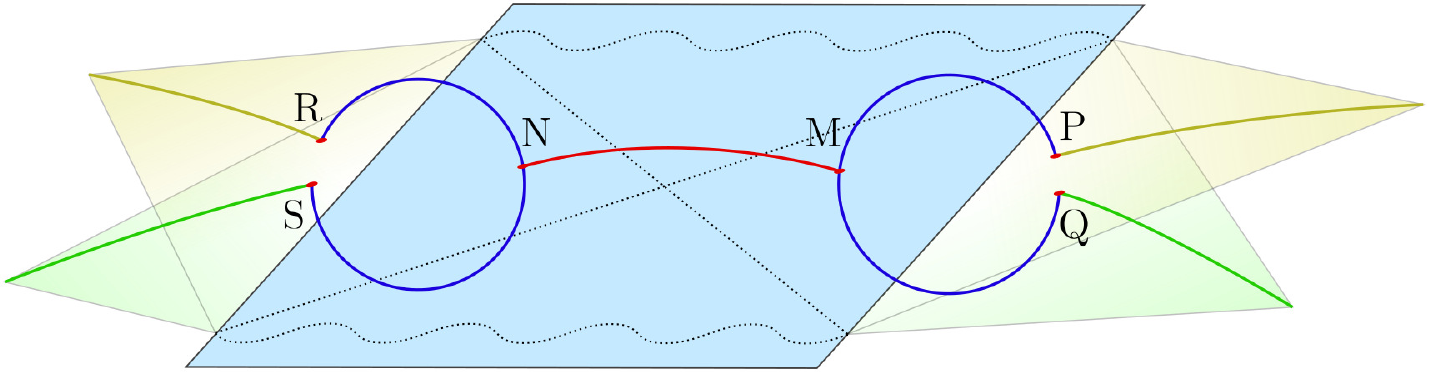}
	\caption{Schematics for the semi-infinite subsystem $A$ in phase-II where we observe the island region $NM$ on the JT brane in the effective $2d$ description.}
	\label{fig:semi_infiniteT-island}
\end{figure}

\subsection*{Doubly holographic description}
This subsection describes the doubly holographic computation of the entanglement entropy for the semi-infinite intervals in the dual CFT$_2 ^{\text{I,II}}$s at a finite temperature as depicted in \cref{fig:semi_infiniteT-island}. In particular, we compute the lengths of the RT surfaces homologous to the semi-infinite intervals. Similar to the previous case, we perform the computation in the planar coordinates\footnote{Note that it is convenient to work with the $(x,t)$ coordinates in this case where we have a planar brane profile.} where the endpoints of the intervals are described by $({x}_0, {t}_0)_\text{I,II}$ (and similarly for the TFD copies) in the dual CFT$_2 ^{\text{I,II}}$s, whereas the island point on the EOW brane is located at ($y$, ${t}_y$). The length of the RT surface may now be obtained to be
\begin{align}\label{sbulk0}
	d =&{2L_\text{I}} \log \left[\frac{\left(x_0+y \sin \psi _\text{I}\right)^2+\left(t_0-t_y\right)^2+\left(y \cos \psi _\text{I}\right)^2}{\epsilon ~y \cos \psi _\text{I}}\right]\nonumber\\
	&+{2L_\text{II}} \log \left[\frac{\left(x_0+y \sin \psi _\text{II}\right)^2+\left(t_0-t_y \right)^2+\left(y \cos \psi _\text{II}\right)^2}{\epsilon ~y \cos \psi _\text{II}}\right],
\end{align}
where the factor 2 arises from the symmetry of the TFD state. After introducing transverse fluctuations on the EOW brane and identifying the dilaton, the entanglement entropy may be expressed as
\begin{align}\label{Sbulk2}
	S_{\rm bulk} = \frac{2\Phi_r}{y} &+ \frac{L_\text{I}}{2G_N}  \log \left(  \frac{x_0^2+2 x_0 y \sin \psi_\text{I}+y^2}{ \epsilon\,y\,\cos \psi_\text{I}}\right) \notag\\
	&+ \frac{L_\text{II}}{2G_N} \log \left(  \frac{x_0^2+2 x_0 y \sin \psi_\text{II}+y^2}{\epsilon \, y \,\cos \psi_\text{II}}\right)\, ,
\end{align}
where extremization over $t_y$ has been performed to set $t_y=t_0$. The location of the island $y$ may now be obtained by extremizing the above to obtain \cref{ext_zeroT_semi} which may be transformed using the maps in \cref{Barnados,cyl-map} to obtain the corresponding extremization condition for the present scenario. Finally it may be observed that in the large tension limit the corresponding results match in the effective 2d theory.

\subsection{Page curve}
\begin{figure}[ht]
	\centering
	\includegraphics[scale=.6]{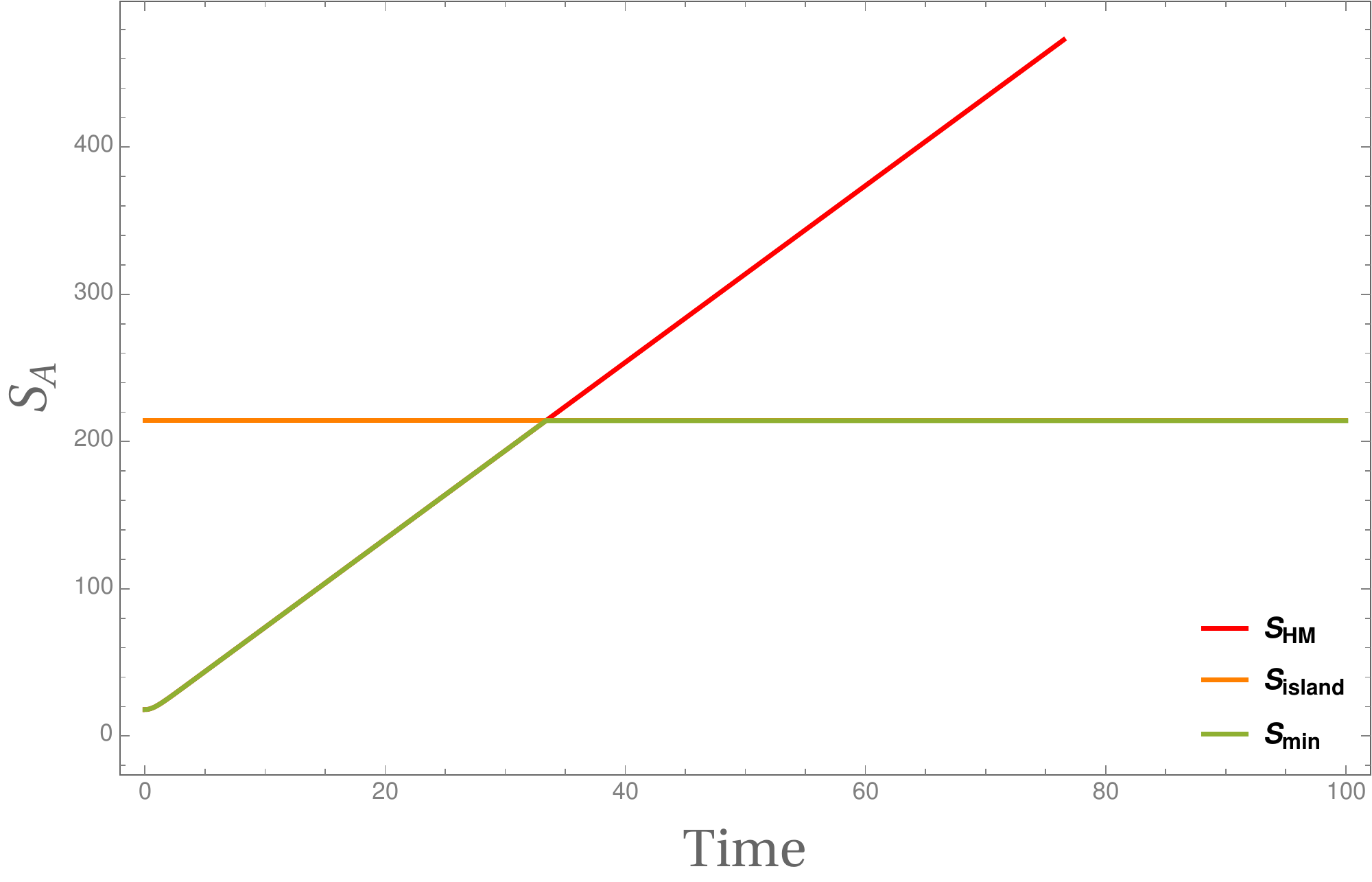}
	\caption{Page curve for semi-finite subsystem in the dual CFT$^\text{I,II}$. Here we have chosen $c_1=6, c_2=12, u_0=10, \beta =2 \pi, \epsilon =0.1$.}
	\label{pagecurve1}
\end{figure}

We now plot the Page curve for the entanglement entropy for the semi-infinite subsystem under consideration in the dual CFT$_2 ^{\text{I,II}}$s at a finite temperature in \cref{pagecurve1}. We observe that, similar to the conventional scenarios with a single CFT$_2$ bath, initially \hyperref[sec:FT-I]{phase-I} is dominant with a monotonically increasing entanglement entropy and finally the island saddle for \hyperref[sec:FT-II]{phase-II} takes over when the entropy gets saturated to a constant value. This is expected as the presence of the additional bath does not affect the radiation process of the JT black hole. It just provides an additional reservoir for the Hawking radiations to be collected. 

\section{Islands and replica wormholes : gravity coupled with two baths} \label{sec:replica}
In this section, we investigate the replica wormhole saddle for the gravitational path integral and reproduce the location of the conical singularity and the entanglement entropy. We first perform the analysis for the effective lower dimensional model obtained from the AdS/ICFT setup by integrating out the bulk degrees of freedom, namely the ``brane+bath'' picture with topological gravity on the AdS$_2$ brane. Later on, we will include JT gravity on the brane and obtain the location of the island and the corresponding fine-grained entropy.

\begin{figure}[ht]
	\centering
	\includegraphics[scale=0.9]{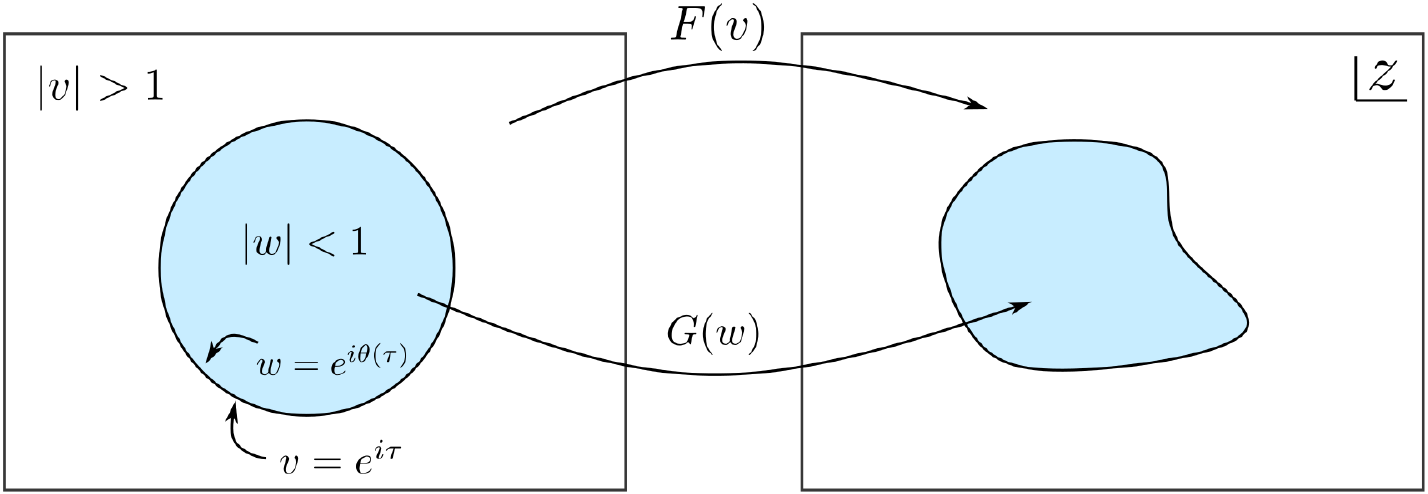}
	\caption{The conventional conformal welding problem \cite{Almheiri:2019qdq}. The two disks $|v|>1$ and $|w|<1$ are glued in terms of the boundary mode $\theta(\tau)$, where $w=e^{i\theta(\tau)}$ and $v=e^{i\tau}$.}
	\label{fig:Welding0}
\end{figure}

The procedure for obtaining the replica wormhole solutions from the boundary curve in two-dimensional gravity coupled to flat bath requires solving the so called conformal welding problem \cite{Almheiri:2019qdq,Goto:2020wnk}. The schematics of the welding problem is sketched in \cref{fig:Welding0}. Essentially, the problem consists in finding a new Riemann surface out of two regions inside and outside of a disk which are described by different coordinate patches. Consider the regions parametrized by $|w|<1$ and $|v|>1$ which are glued together along their boundaries at $|v|=|w|=1$, where the complex coordinates are described by
\begin{align}
	v=e^y=e^{\sigma+i\tau}~~,~~w=e^{\gamma+i\theta}\,.\label{bath-bulk-complexcoord}
\end{align}
It is, in general, impossible to extend the coordinates $w$ or $v$ holomorphically beyond the respective boundary circles. However, by virtue of the Riemann mapping theorem, one can find two holomorphic functions $F$ and $G$ to establish another coordinate system $z$ on a new Riemann surface such that the regions $|w|<1$ and $|v|>1$ are holomorphically mapped to the coordinate $z$. In other words, one requires 
\begin{align}
	z=G(w)~~,~~&\text{for} ~|w|<1\notag\\
	z=F(v)~~,~~&\text{for} ~|v|>1\\
	G\left(e^{i\theta(\tau)}\right)=F\left(e^{i\tau}\right)~~,~~&\text{for} ~|v|=|w|=1\,.\notag
\end{align}
The problem of finding holomorphic $F(v)$ and $G(w)$ given the boundary mode $\theta(\tau)$ is termed the \textit{conformal welding} problem. In the case of two dimensional gravity on a AdS$_2$ manifold coupled to a flat CFT$_2$ bath such a welding issue arises naturally \cite{Almheiri:2019qdq, Goto:2020wnk}. In the presence of dynamical gravity, the entanglement entropy for a subsystem is computed through the Lewkowycz-Maldacena procedure by considering an $n$-fold cover of the original manifold $\mathcal{M}$ \cite{Lewkowycz:2013nqa}. For a replica symmetric saddle $\mathcal{M}_n$ to the gravitational path integral, it is convenient to quotient by the $\mathbb{Z}_n$ replica symmetry and consider a single manifold $\tilde{\mathcal{M}}_n=\mathcal{M}_n/\mathbb{Z}_n$. The orbifold $\tilde{\mathcal{M}}_n$ essentially describes a disk with conical singularities at which twist operators for the conformal matter theory are inserted. The metric on the interior manifold $\tilde{\mathcal{M}}_n$ may be described by a complex coordinate $w$ as follows:
\begin{align}
	\text{d}s^2=e^{2\rho(w,\bar{w})}\text{d}w\text{d}\bar{w}~~,~~ \text{for} |w|<1\,.
\end{align}
In a finite temperature configuration with $\tau\sim \tau+2\pi$,
in order to join the metric of the quotient manifold of the gravitating region to the flat space outside described by the exterior coordinates $v=e^y$, it is required to solve the conformal welding problem discussed above. In this case, the boundary mode $\theta(\tau)$ plays the role of the reparametrization mode in two-dimensional gravity \cite{Almheiri:2019qdq}.
\subsection{Replica wormholes from AdS/ICFT}
\begin{figure}[ht]
	\centering
	\includegraphics[scale=0.7]{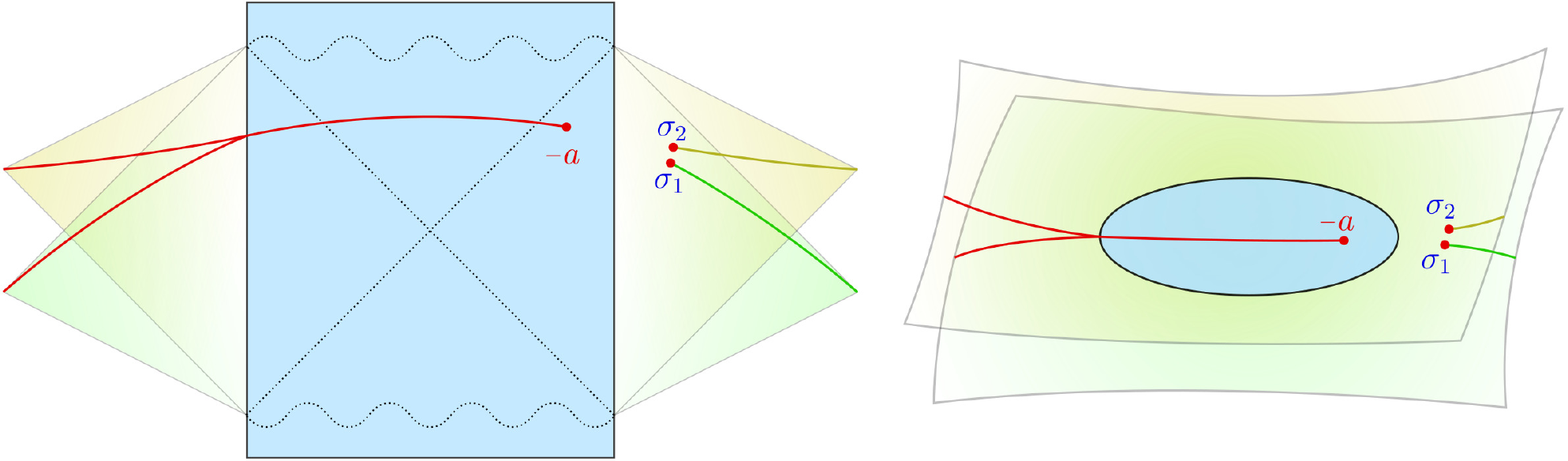}
	\caption{The single sided configuration in the Lorentzian signature (left) and in the Euclidean signature (right). }
	\label{fig:ICFT-RW}
\end{figure}

In this subsection, we focus on the replica wormhole solutions in the framework of AdS/ICFT discussed in \cite{Anous:2022wqh} and briefly reviewed in \cref{sec:review}. In the effective lower dimensional scenario obtained from integrating out the bulk spacetimes on either side of the brane $\sigma$, we have two flat baths attached to the gravitational region on the EOW brane $\Sigma$ which has a weakly gravitating metric in the large tension limit. There are two CFTs along the flat half lines which extends to the gravitating region where they interact via the weakly fluctuating metric. The schematics of the setup is sketched in \cref{fig:ICFT-RW}.

As discussed earlier, for a quantum field theory coupled to dynamical gravity on a hybrid manifold, the replica trick to compute the entanglement entropy for a subsystem involves a replication of the original manifold in the replica index $n$. The normalized partition function $\mathbf{Z}_n$ on this replica manifold then computes the entanglement entropy as follows \cite{Lewkowycz:2013nqa,Almheiri:2019qdq}
\begin{align}
	S=-\partial_n\left(\frac{\log\mathbf{Z}_n}{n}\right)\Bigg|_{n=1}\,.
\end{align}
The partition function for the gravity region concerns a gravitational path integral which may be solved in the saddle-point approximation in the semi-classical regime by specifying appropriate boundary conditions. These saddles may be characterized by the nature of gluing of the individual replica copies. In particular, two specific choices will be of importance for our purposes, namely the Hawking saddle where the $n$-copies of the bath(s) are glued cyclically while gravity is filled in each copy individually, and the replica wormhole saddle in which along with the copies of the bath, gravitational regions are dynamically glued together. In these replica wormhole saddles, upon quotienting via the replica symmetry $\mathbb{Z}_n$, additional conical singularities dynamically appear at the fixed points of the replica symmetry in the orbifold theory.

The gravitational action on the orbifold $\tilde{\Sigma}_n$ obtained by quotienting the replicated EOW brane $\Sigma_n$ is given by
\begin{align}
	-\frac{1}{n}I_{\text{grav}}\big[\tilde{\Sigma}_n\big]=\sum_{k=\text{I} , \text{II}}\frac{L_k}{32\pi G_N}\int_{\Sigma}\text{d}^2 y \sqrt{-\tilde{h}}\,\bigg[R^{(2)}-R^{(2)}&\log\left(-\frac{L^2_k}{2}R^{(2)}\right)\bigg]\notag\\
	-&\left(1-\frac{1}{n}\right)\sum_i S(w_i)\,,
\end{align}
where $S(w_i)$ denotes the contributions from the dynamical conical singularities. In our case, this is just a constant given in \cref{AreaPoly} with a vanishing dilaton term.

\begin{figure}[ht]
	\centering
	\includegraphics[scale=0.8]{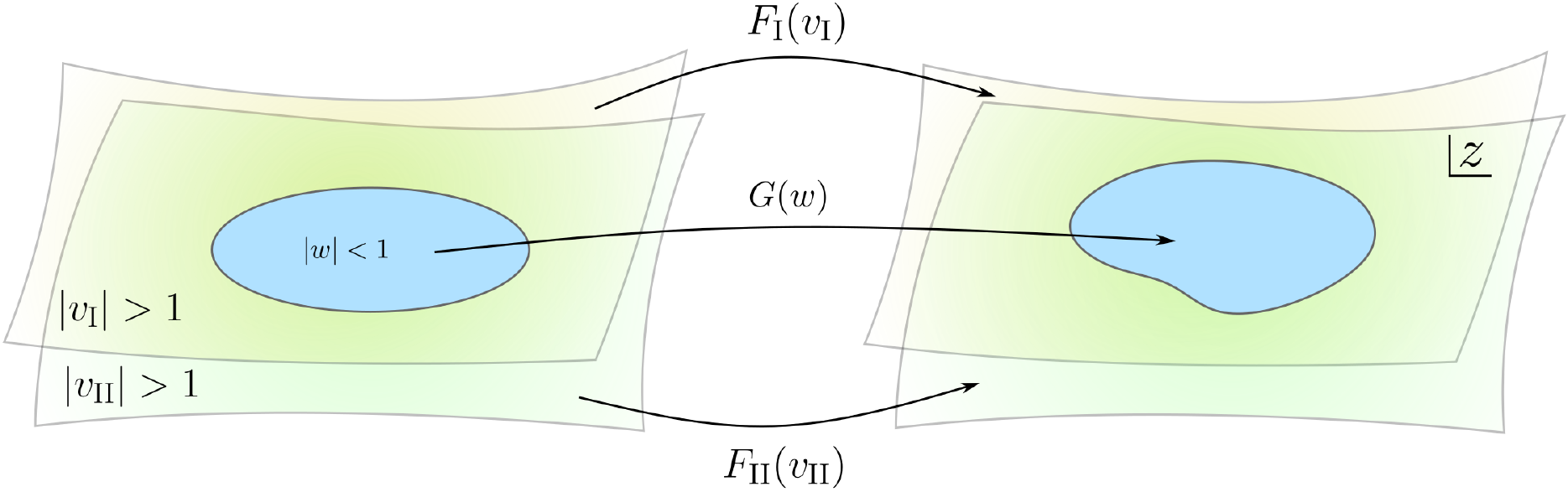}
	\caption{Conformal welding problem for our setup with two CFT$_2$ baths defined in the regions $|v_{\text{I}}|>1$ and $|v_{\text{II}}|>1$ which are coupled to gravity inside the circle $|w|<1$. }
	\label{fig:Welding}
\end{figure}

We choose the complex coordinate $w$ to describe the gravity region inside the disk $|w|=1$. Furthermore, the baths outside the disk are described by the complex coordinates $v_k\,,(k=\text{{I,II}})$, in the spirit of \cref{bath-bulk-complexcoord}. Then the conformal welding problem sketched in \cref{fig:Welding} is reduced to the determination of the appropriate boundary mode $\theta(\tau)$. We consider two semi-infinite intervals in the bath CFT$_2^\text{I,II}$s as $[\sigma_1,\infty]_\text{I}$ and $[\sigma_2,\infty]_\text{II}$ and in the replica manifold twist operators are placed at the locations $v_\text{I}=e^{\sigma_1}$ and $v_\text{II}=e^{\sigma_2}$. Note that for the replica wormhole saddle, a dynamical conical singularity also appears at $w=e^{-a}$.

To proceed, we now require the energy flux equation at the interface of the gravitational region and the bath CFTs. The variation of the gravitational action with respect to the boundary mode is vanishing
\begin{align}
	-\frac{1}{n}\,\delta I_{\text{grav}}=0\,.
\end{align}

On the other hand, the variation of the matter partition function $\mathbf{Z}_{\text{mat}}$ with respect to the boundary mode leads to the following expression \cite{Almheiri:2019qdq}
\begin{align}
	\delta\log\mathbf{Z}_{\text{mat}}=i\int\text{d}\tau\sum_{k=\text{I} , \text{II}}\left(T^{(k)}_{yy}-T^{(k)}_{\bar{y}\bar{y}}\right)\frac{\delta\theta(\tau)}{\theta'(\tau)}
\end{align}
Utilizing the above equations, the energy flux condition at the boundary may be expressed as follows
\begin{align}
	i\left[T^{(\text{I})}_{yy}(i\tau)-T^{(\text{I})}_{\bar{y}\bar{y}}(-i\tau)\right]+i\left[T^{(\text{II})}_{yy}(i\tau)-T^{(\text{II})}_{\bar{y}\bar{y}}(-i\tau)\right]=0\,.\label{E-Flux-ICFT}
\end{align}
Under the conformal map $y\to z=F_k(v_k)$, the energy momentum tensor transforms as
\begin{align}
	T^{(k)}_{yy}(i\tau)\to e^{2y}\left[\left(\frac{dF_{k}(e^y)}{dv_k}\right)^2 T^{(k)}_{zz}-\frac{c_k}{24\pi}\left\{F_{k}(e^y),v_k\right\}\right]~~,~~k=\text{I} , \text{II}
\end{align}
In the replicated geometry, the uniformization map for the conical singularities is given by $z\to \tilde{z}=z^{1/n}$ such that $T^{({k})}_{\tilde{z}\tilde{z}}=0$ and the energy-momentum tensors for the two CFTs in the $z$-plane is given by
\begin{align}
	T^{({k})}_{zz}=-\frac{c_k}{24\pi}\left(1-\frac{1}{n^2}\right)\frac{1}{z^2}\,.
\end{align}
Therefore, the energy-flux condition in \cref{E-Flux-ICFT} reduces to
\begin{align}
	0=\sum_{k=\text{I} , \text{II}}i e^{2i\tau}c_k\left[\frac{1}{2}\left(1-\frac{1}{n^2}\right)\left(\frac{{F_{k}}'(e^{i\tau})}{F_{k}(e^{i\tau})}\right)^2+\left\{F_{k}(e^{i\tau}),e^{i\tau}\right\}\right]+\text{c.c.}\,.\label{EFlux1}
\end{align}
Since the maps $F_{k}$ depend on the gluing function $\theta(\tau)$, the above equation is in general hard to solve. However, one may solve it near $n=1$ as described below. 

For $n=1$, the first term in the parenthesis of \cref{EFlux1} vanishes and the welding is trivial. Therefore, we may conclude that the maps $F_{k}$ are well approximated near $n=1$ by M\"obius transformations of the form
\begin{align}
	z=F_{k}(v_k)=\frac{v_k-\mathcal{A}}{\mathcal{B}_k-v_k}~~,~~ \mathcal{A} = e^{-a}~ , ~\mathcal{B}_\text{I} = e^{\sigma_1}~ , ~\mathcal{B}_\text{II} = e^{\sigma_2} \label{Fzero}
\end{align}
It is straightforward to verify that these functions indeed map the branch points at $-a$ and $\sigma_{1,2}$ to $z=0$ and $z=\infty$ respectively. Therefore, the energy flux condition becomes
\begin{align}
	c_{\text{I}}\mathcal{F}_{\text{I}}+c_{\text{II}}\mathcal{F}_{\text{II}}=0\,,
\end{align}
where 
\begin{align}
	\mathcal{F}_{k}=i e^{2i\tau}\left(\frac{{F_k}'\left(e^{i\tau}\right)}{F_k\left(e^{i\tau}\right)}\right)^2+\text{c.c.}=i e^{2i\tau}\frac{(\mathcal{A}-\mathcal{B}_k)^2}{\left(e^{-i\tau}-\mathcal{A}\right)^2\left(e^{i\tau}-\mathcal{B}_k\right)^2}+\text{c.c.}
\end{align}
Now, performing a Fourier transformation in the above equation (restoring the temperature $\beta$), the expression for the $k=1$ mode reads
\begin{align}
	0&=\int_{0}^{\beta}\, \text{d}\tau\, e^{-\frac{2\pi i\tau}{\beta}}\left(c_{\text{I}}\mathcal{F}_{\text{I}}+c_{\text{II}}\mathcal{F}_{\text{II}}\right)\notag\\
	&=c_{\text{I}}\left(\frac{\sinh\left[\frac{\pi(a-\sigma_1)}{\beta}\right]}{\sinh\left[\frac{\pi(a+\sigma_1)}{\beta}\right]}\right)+c_{\text{II}}\left(\frac{\sinh\left[\frac{\pi(a-\sigma_2)}{\beta}\right]}{\sinh\left[\frac{\pi(a+\sigma_2)}{\beta}\right]}\right)
\end{align}
In order to compare with the quantum extremal surface condition at zero temperature, we now take the $\beta\to\infty$ limit to obtain
\begin{align}
	0=c_{\text{I}}\left(\frac{a-\sigma_1}{a+\sigma_1}\right)+c_{\text{II}}\left(\frac{a-\sigma_2}{a+\sigma_2}\right)\,,
\end{align}
which on solving for $a$ gives
\begin{align}
	a^*=\frac{(c_{\text{I}}-c_{\text{II}})\left(\sigma_1-\sigma _2\right)+\sqrt{4 \left(c_{\text{I}}+c_{\text{II}}\right)^2\sigma_1 \sigma_2+\left(c_{\text{I}}-c_{\text{II}}\right)^2 \left(\sigma_1-\sigma_2\right)^2}}{2 \left(c_{\text{I}}+c_{\text{II}}\right)}\,.
\end{align}
The above expression is identical to the position of the quantum extremal surface obtained through extremizing the generalized entropy in \cite{Anous:2022wqh}.

\subsection{Replica wormholes with JT gravity coupled to two baths}
With JT gravity on the EOW brane $\Sigma$, the energy flux condition at the boundary of the replicated geometry is modified and the conformal welding problem is a bit more involved. The variation of the gravitational action with respect to the boundary mode $\theta(\tau)$ no longer vanishes since in the case of JT gravity $\theta(\tau)$ serves as the ``boundary graviton'' \cite{Almheiri:2019qdq}. The energy flux condition in the presence of JT gravity on the brane is then modified to \cite{Almheiri:2019qdq,Goto:2020wnk,Suzuki:2022xwv}
\begin{align}
	\partial_{\tau}M=	i\left[T^{(\text{I})}_{yy}(i\tau)-T^{(\text{I})}_{\bar{y}\bar{y}}(-i\tau)\right]+i\left[T^{(\text{II})}_{yy}(i\tau)-T^{(\text{II})}_{\bar{y}\bar{y}}(-i\tau)\right]\label{EFluxJT}
\end{align}
where $M$ corresponds to the ADM mass of the gravitational theory which is related to the Schwarzian boundary action.

For the two single intervals $[\sigma_1,\infty]_\text{I}$ and $[\sigma_2,\infty]_\text{II}$ on CFT$_2^\text{I,II}$ baths, a conical singularity appears inside the gravity region on the orbifold theory $\Sigma_n/\mathbb{Z}_n$ at a point $-a$ and we need to consider the subsystems $[-\infty,-a]_\text{I}\cup [\sigma_1,\infty]_\text{I}$ and $[-\infty,-a]_\text{II}\cup [\sigma_2,\infty]_\text{II}$\footnote{Note that the local geometry at the point $-a$ on the replica manifold $\Sigma_n$ is completely smooth.}. Once again, we will work with a finite temperature configuration with $\beta=2\pi$. We may now uniformize the interior conical singularity at $w=\mathcal{A}=e^{-a}$ by utilizing the map
\begin{align}
	\tilde{w}=\left(\frac{w-\mathcal{A}}{1-\mathcal{A}\,w}\right)^{\frac{1}{n}}\,.
\end{align}
In the $\tilde{w}$ coordinates, the gravity region has the usual hyperbolic disk metric \cite{Almheiri:2019qdq}
\begin{align}
	\text{d}s^2_{\text{in}}=\frac{4\text{d}\tilde{w}\text{d}\bar{\tilde{w}}}{\left(1-|\tilde{w}|^2\right)^2}\,.
\end{align}
In these coordinates we may set $\tilde{w}=e^{i\tilde{\theta}}$ at the boundary. Now using the Schwarzian composition rules, we may obtain the ADM mass of the spacetime to be \cite{Almheiri:2019qdq}
\begin{align}
	\frac{\Phi_r}{4\pi}\left\{e^{i\tilde{\theta}},\tau\right\}=\frac{\Phi_r}{4\pi}\left[\left\{e^{i\theta},\tau\right\}+\frac{1}{2}\left(1-\frac{1}{n^2}\right)R(\theta)\right]\,,\label{ADMJT}
\end{align}
where the function $R(\theta)$ contains the information about the branch point $-a$ as follows
\begin{align}
	R(\theta)=-\frac{(1-\mathcal{A}^2)^2(\partial_{\tau}\theta)^2}{|1-\mathcal{A}\,e^{i\theta}|^4}\,.
\end{align}
Now, utilizing \cref{EFlux1,EFluxJT,ADMJT}, the energy flux condition at the boundary becomes
\begin{align}
	-\frac{12\Phi_r}{c}&\partial_{\tau}\left[\left\{e^{i\theta},\tau\right\}+\frac{1}{2}\left(1-\frac{1}{n^2}\right)R(\theta)\right]\notag\\
	&=\sum_{k=\text{I} , \text{II}}i e^{2i\tau}c_k\left[\frac{1}{2}\left(1-\frac{1}{n^2}\right)\left(\frac{{F_{k}}'(e^{i\tau})}{F_{k}(e^{i\tau})}\right)^2+\left\{F_{k}(e^{i\tau}),e^{i\tau}\right\}\right]+\text{c.c.}\label{EFluxJT1}
\end{align}
The above relation is quite complicated as the map $F$ depends implicitly on the gluing function $\theta(\tau)$. Nevertheless, as earlier, we may solve it near $n\sim 1$ as follows.
Near $n\sim 1$, we may expand the boundary mode $\theta(\tau)$ as follows \cite{Almheiri:2019qdq}
\begin{align}
	e^{i\theta(\tau)}=e^{i\tau}\left[1+i\delta\theta(\tau)\right]\,,
\end{align}
where $\delta\theta(\tau)$ is of order $(n-1)$. Next we use the following relation \cite{Almheiri:2019qdq}
\begin{align}
	e^{2i\tau}\left\{F_k,e^{i\tau}\right\}=-\frac{1}{2}(1+H)(\delta\theta'''+\delta\theta')\,,
\end{align}
where $H$ is the Hilbert transform\footnote{The Hilbert transform is defined through the action 
	\begin{align}
		H \cdot e^{im\tau}=-\text{sgn}(m)e^{im\tau}~~,~~H\cdot 1=0\,.
\end{align}} which projects out the negative frequency modes of $\delta\theta$. Note that, except for in the Schwarzian term $\left\{F_k,e^{i\tau}\right\}$, the functions $F_k$ appear with a factor of $(n-1)$ in \cref{EFluxJT1} and we may keep only up to the zeroth order solutions given in \cref{Fzero}.
Restoring the temperature dependence utilizing the scaling $\Phi_r\to \frac{2\pi\Phi_r}{\beta}$ and Fourier transforming to the $k$ basis, the $k=1$ mode requires
\begin{align}
	\int_{0}^{\beta}\text{d}\tau \,e^{-\frac{2\pi i\tau}{\beta}}\left[c_{\text{I}}\mathcal{F}_{\text{I}}+c_{\text{II}}\mathcal{F}_{\text{II}}-\frac{{24\pi\Phi_r}}{\beta}\partial_{\tau}R(\tau)\right]=0
\end{align}
which leads to the condition
\begin{align}
	c_{\text{I}}\left(\frac{\sinh\left[\frac{\pi(a-\sigma_1)}{\beta}\right]}{\sinh\left[\frac{\pi(a+\sigma_1)}{\beta}\right]}\right)+c_{\text{II}}\left(\frac{\sinh\left[\frac{\pi(a-\sigma_2)}{\beta}\right]}{\sinh\left[\frac{\pi(a+\sigma_2)}{\beta}\right]}\right)=\frac{{12\pi\Phi_r}}{\beta}\text{csch}\left(\frac{2\pi a}{\beta}\right)
\end{align}
In the $\beta\to \infty$ limit, this reduces to a cubic equation in $a$ as follows
\begin{align}
	c_{\text{I}}\left(\frac{a-\sigma_1}{a+\sigma_1}\right)+c_{\text{II}}\left(\frac{a-\sigma_2}{a+\sigma_2}\right)=\frac{6\Phi_r}{a}
\end{align}
The above equation is easily solved for $a$ but the solutions are not quite illuminating. Instead, we take the simplifying limit $\sigma_1=\sigma_2=\sigma$ to get the quadratic equation
\begin{align}
	a (a-\sigma )-\frac{6 \Phi_r}{\left(c_{\text{I}}+c_{\text{II}}\right)} (a+\sigma)=0\,.
\end{align} 
Solving for the position of the conical singularity $a$, we obtain
\begin{align}
	a^*=\frac{\left(c_{\text{I}}+c_{\text{II}}\right) \sigma +6 \Phi_r+\sqrt{\left(\left(c_{\text{I}}+c_{\text{II}}\right) \sigma +6 \Phi_r\right)^2+24 \left(c_{\text{I}}+c_{\text{II}}\right) \sigma \Phi_r}}{2 \left(c_{\text{I}}+c_{\text{II}}\right)}
\end{align}
which is identical to the position of the QES obtained in \cref{sec:single-crossing}.

\section{Summary and discussion}\label{sec:summary}

In this article, we have investigated the entanglement structure of various bipartite states in a hybrid manifold where a JT gravity is coupled to two non-gravitating CFT$_2$ baths. To this end, we first construct this hybrid theory through a dimensional reduction of a $3d$ geometry. The $3d$ geometry is comprised of a fluctuating EOW brane acting as an interface between two distinct AdS$_3$ geometries. Performing a partial Randall-Sundrum reduction in the neighbourhood of the fluctuating brane results in JT gravity on the EOW brane. Furthermore, utilizing the usual AdS/CFT correspondence on the remaining wedges of the two AdS$_3$ geometries leads to two non-gravitating CFT baths on two half lines. In the limit of large brane tension, we obtain the $2d$ effective theory of JT gravity coupled to conformal matter on the hybrid ``brane+baths'' manifolds.

Furthermore, we have provided a prescription for computing the generalized R\'enyi entropy for a subsystem in this hybrid manifold. In particular, for this scenario where the JT gravity is coupled to two CFT$_2$ baths, the dominant replica wormhole saddle is modified to provide two independent mechanisms to obtain an island region. Other than the conventional origin of the island region where the degrees of freedom for the CFT in the gravitational region is shared by bath CFTs, we also observe cases where island region is captured for CFT$^\text{I}$ even though no bath degrees of freedom is considered. We have called such regions as the induced islands, as the subsystem purely in CFT$^\text{II}$ induces an island region even for CFT$^\text{I}$. 
In the doubly holographic perspective this phenomena corresponds to the double-crossing geodesic where the RT surface crosses from AdS$^\text{II}$ to AdS$^\text{I}$ and returns to AdS$^\text{II}$. 

Subsequently, we obtain the entanglement entropy for subsystems comprised of semi-infinite and finite intervals in CFT$_2^\text{I,II}$ coupled to extremal as well as eternal JT black holes. We perform computations from the effective $2d$ perspective using the generalized entanglement entropy formula and find agreement with the doubly holographic computation in the large tension limit of the EOW brane for all the cases. We also plot Page curves for the different configurations of the subsystems and observe transitions between different phases of the entanglement entropy. 

We have also performed the so called conformal welding problem for the replica wormhole saddle in the effective ``brane+bath'' scenario and obtained the location of the island for semi-infinite subsystems in the baths. To this end, we begin with the lower dimensional effective picture obtained from the AdS/ICFT setup discussed in \cite{Anous:2022wqh} and reproduce the QES result. Subsequently, this is extended to the case with JT gravity on the EOW brane which substantiate the island computations for the corresponding configuration.

There are several future directions to explore. For finite intervals in the baths coupled to JT gravity, the location of the islands may be obtained through the conformal welding problem with the replica wormhole by extending of the analysis in \cite{Goto:2020wnk}. It will be interesting to explore the nature of mixed state entanglement in Hawking radiation from the JT black hole via different entanglement and correlation measures such as the reflected entropy \cite{Dutta:2019gen}, the entanglement negativity \cite{Calabrese:2012nk}, the entanglement of purification \cite{Takayanagi:2017knl} and the balanced partial entanglement \cite{Wen:2021qgx}. Furthermore, our setup can be extended to include holographic models of interface CFTs which involve two interface branes separating three bulk regions \cite{Baig:2022cnb}. A partial dimensional reduction on different bulk wedges would result in two fluctuating JT branes with black holes which interact through the CFT$_2$ baths on a hybrid seagull-like geometry. This provides yet another exotic model of Hawking radiation which may lead to new insights for the information loss problem.

\acknowledgments The work of GS is partially supported by the Dr Jagmohan Garg Chair Professor position at the Indian Institute of Technology, Kanpur.

\bibliographystyle{JHEP}

\bibliography{reference}

\providecommand{\href}[2]{#2}\begingroup\raggedright\begin{thebibliography}{10}

\bibitem{Penington:2019npb}
G.~Penington, \emph{{Entanglement Wedge Reconstruction and the Information
  Paradox}}, \href{https://doi.org/10.1007/JHEP09(2020)002}{\emph{JHEP}
  {\bfseries 09} (2020) 002}
  [\href{https://arxiv.org/abs/1905.08255}{{\ttfamily 1905.08255}}].

\bibitem{Penington:2019kki}
G.~Penington, S.~H. Shenker, D.~Stanford and Z.~Yang, \emph{{Replica wormholes
  and the black hole interior}},
  \href{https://doi.org/10.1007/JHEP03(2022)205}{\emph{JHEP} {\bfseries 03}
  (2022) 205} [\href{https://arxiv.org/abs/1911.11977}{{\ttfamily
  1911.11977}}].

\bibitem{Almheiri:2019hni}
A.~Almheiri, R.~Mahajan, J.~Maldacena and Y.~Zhao, \emph{{The Page curve of
  Hawking radiation from semiclassical geometry}},
  \href{https://doi.org/10.1007/JHEP03(2020)149}{\emph{JHEP} {\bfseries 03}
  (2020) 149} [\href{https://arxiv.org/abs/1908.10996}{{\ttfamily
  1908.10996}}].

\bibitem{Almheiri:2019qdq}
A.~Almheiri, T.~Hartman, J.~Maldacena, E.~Shaghoulian and A.~Tajdini,
  \emph{{Replica Wormholes and the Entropy of Hawking Radiation}},
  \href{https://doi.org/10.1007/JHEP05(2020)013}{\emph{JHEP} {\bfseries 05}
  (2020) 013} [\href{https://arxiv.org/abs/1911.12333}{{\ttfamily
  1911.12333}}].

\bibitem{Almheiri:2019yqk}
A.~Almheiri, R.~Mahajan and J.~Maldacena, \emph{{Islands outside the horizon}},
   \href{https://arxiv.org/abs/1910.11077}{{\ttfamily 1910.11077}}.

\bibitem{Page:1993wv}
D.~N. Page, \emph{{Information in black hole radiation}},
  \href{https://doi.org/10.1103/PhysRevLett.71.3743}{\emph{Phys. Rev. Lett.}
  {\bfseries 71} (1993) 3743}
  [\href{https://arxiv.org/abs/hep-th/9306083}{{\ttfamily hep-th/9306083}}].

\bibitem{Page:1993df}
D.~N. Page, \emph{{Average entropy of a subsystem}},
  \href{https://doi.org/10.1103/PhysRevLett.71.1291}{\emph{Phys. Rev. Lett.}
  {\bfseries 71} (1993) 1291}
  [\href{https://arxiv.org/abs/gr-qc/9305007}{{\ttfamily gr-qc/9305007}}].

\bibitem{Page:2013dx}
D.~N. Page, \emph{{Time Dependence of Hawking Radiation Entropy}},
  \href{https://doi.org/10.1088/1475-7516/2013/09/028}{\emph{JCAP} {\bfseries
  09} (2013) 028} [\href{https://arxiv.org/abs/1301.4995}{{\ttfamily
  1301.4995}}].

\bibitem{Sully:2020pza}
J.~Sully, M.~V. Raamsdonk and D.~Wakeham, \emph{{BCFT entanglement entropy at
  large central charge and the black hole interior}},
  \href{https://doi.org/10.1007/JHEP03(2021)167}{\emph{JHEP} {\bfseries 03}
  (2021) 167} [\href{https://arxiv.org/abs/2004.13088}{{\ttfamily
  2004.13088}}].

\bibitem{Rozali:2019day}
M.~Rozali, J.~Sully, M.~Van~Raamsdonk, C.~Waddell and D.~Wakeham,
  \emph{{Information radiation in BCFT models of black holes}},
  \href{https://doi.org/10.1007/JHEP05(2020)004}{\emph{JHEP} {\bfseries 05}
  (2020) 004} [\href{https://arxiv.org/abs/1910.12836}{{\ttfamily
  1910.12836}}].

\bibitem{Chen:2020uac}
H.~Z. Chen, R.~C. Myers, D.~Neuenfeld, I.~A. Reyes and J.~Sandor,
  \emph{{Quantum Extremal Islands Made Easy, Part I: Entanglement on the
  Brane}}, \href{https://doi.org/10.1007/JHEP10(2020)166}{\emph{JHEP}
  {\bfseries 10} (2020) 166}
  [\href{https://arxiv.org/abs/2006.04851}{{\ttfamily 2006.04851}}].

\bibitem{Chen:2020hmv}
H.~Z. Chen, R.~C. Myers, D.~Neuenfeld, I.~A. Reyes and J.~Sandor,
  \emph{{Quantum Extremal Islands Made Easy, Part II: Black Holes on the
  Brane}}, \href{https://doi.org/10.1007/JHEP12(2020)025}{\emph{JHEP}
  {\bfseries 12} (2020) 025}
  [\href{https://arxiv.org/abs/2010.00018}{{\ttfamily 2010.00018}}].

\bibitem{Grimaldi:2022suv}
G.~Grimaldi, J.~Hernandez and R.~C. Myers, \emph{{Quantum extremal islands made
  easy. Part IV. Massive black holes on the brane}},
  \href{https://doi.org/10.1007/JHEP03(2022)136}{\emph{JHEP} {\bfseries 03}
  (2022) 136} [\href{https://arxiv.org/abs/2202.00679}{{\ttfamily
  2202.00679}}].

\bibitem{Suzuki:2022xwv}
K.~Suzuki and T.~Takayanagi, \emph{{BCFT and Islands in two dimensions}},
  \href{https://doi.org/10.1007/JHEP06(2022)095}{\emph{JHEP} {\bfseries 06}
  (2022) 095} [\href{https://arxiv.org/abs/2202.08462}{{\ttfamily
  2202.08462}}].

\bibitem{Deng:2020ent}
F.~Deng, J.~Chu and Y.~Zhou, \emph{{Defect extremal surface as the holographic
  counterpart of Island formula}},
  \href{https://doi.org/10.1007/JHEP03(2021)008}{\emph{JHEP} {\bfseries 03}
  (2021) 008} [\href{https://arxiv.org/abs/2012.07612}{{\ttfamily
  2012.07612}}].

\bibitem{Chu:2021gdb}
J.~Chu, F.~Deng and Y.~Zhou, \emph{{Page curve from defect extremal surface and
  island in higher dimensions}},
  \href{https://doi.org/10.1007/JHEP10(2021)149}{\emph{JHEP} {\bfseries 10}
  (2021) 149} [\href{https://arxiv.org/abs/2105.09106}{{\ttfamily
  2105.09106}}].

\bibitem{Takayanagi:2011zk}
T.~Takayanagi, \emph{{Holographic Dual of BCFT}},
  \href{https://doi.org/10.1103/PhysRevLett.107.101602}{\emph{Phys. Rev. Lett.}
  {\bfseries 107} (2011) 101602}
  [\href{https://arxiv.org/abs/1105.5165}{{\ttfamily 1105.5165}}].

\bibitem{Fujita:2011fp}
M.~Fujita, T.~Takayanagi and E.~Tonni, \emph{{Aspects of AdS/BCFT}},
  \href{https://doi.org/10.1007/JHEP11(2011)043}{\emph{JHEP} {\bfseries 11}
  (2011) 043} [\href{https://arxiv.org/abs/1108.5152}{{\ttfamily 1108.5152}}].

\bibitem{Kastikainen:2021ybu}
J.~Kastikainen and S.~Shashi, \emph{{Structure of holographic BCFT correlators
  from geodesics}},
  \href{https://doi.org/10.1103/PhysRevD.105.046007}{\emph{Phys. Rev. D}
  {\bfseries 105} (2022) 046007}
  [\href{https://arxiv.org/abs/2109.00079}{{\ttfamily 2109.00079}}].

\bibitem{Li:2021dmf}
T.~Li, M.-K. Yuan and Y.~Zhou, \emph{{Defect extremal surface for reflected
  entropy}}, \href{https://doi.org/10.1007/JHEP01(2022)018}{\emph{JHEP}
  {\bfseries 01} (2022) 018}
  [\href{https://arxiv.org/abs/2108.08544}{{\ttfamily 2108.08544}}].

\bibitem{Basu:2022reu}
D.~Basu, H.~Parihar, V.~Raj and G.~Sengupta, \emph{{Defect extremal surfaces
  for entanglement negativity}},
  \href{https://arxiv.org/abs/2205.07905}{{\ttfamily 2205.07905}}.

\bibitem{Shao:2022gpg}
Y.~Shao, M.-K. Yuan and Y.~Zhou, \emph{{Entanglement Negativity and Defect
  Extremal Surface}},  \href{https://arxiv.org/abs/2206.05951}{{\ttfamily
  2206.05951}}.

\bibitem{Lu:2022cgq}
Y.~Lu and J.~Lin, \emph{{The Markov gap in the presence of islands}},
  \href{https://doi.org/10.1007/JHEP03(2023)043}{\emph{JHEP} {\bfseries 03}
  (2023) 043} [\href{https://arxiv.org/abs/2211.06886}{{\ttfamily
  2211.06886}}].

\bibitem{Jackiw:1984je}
R.~Jackiw, \emph{{Lower Dimensional Gravity}},
  \href{https://doi.org/10.1016/0550-3213(85)90448-1}{\emph{Nucl. Phys. B}
  {\bfseries 252} (1985) 343}.

\bibitem{Teitelboim:1983ux}
C.~Teitelboim, \emph{{Gravitation and Hamiltonian Structure in Two Space-Time
  Dimensions}}, \href{https://doi.org/10.1016/0370-2693(83)90012-6}{\emph{Phys.
  Lett. B} {\bfseries 126} (1983) 41}.

\bibitem{Deng:2022yll}
F.~Deng, Y.-S. An and Y.~Zhou, \emph{{JT gravity from partial reduction and
  defect extremal surface}},
  \href{https://doi.org/10.1007/JHEP02(2023)219}{\emph{JHEP} {\bfseries 02}
  (2023) 219} [\href{https://arxiv.org/abs/2206.09609}{{\ttfamily
  2206.09609}}].

\bibitem{Geng:2022slq}
H.~Geng, A.~Karch, C.~Perez-Pardavila, S.~Raju, L.~Randall, M.~Riojas et~al.,
  \emph{{Jackiw-Teitelboim Gravity from the Karch-Randall Braneworld}},
  \href{https://doi.org/10.1103/PhysRevLett.129.231601}{\emph{Phys. Rev. Lett.}
  {\bfseries 129} (2022) 231601}
  [\href{https://arxiv.org/abs/2206.04695}{{\ttfamily 2206.04695}}].

\bibitem{Geng:2022tfc}
H.~Geng, \emph{{Aspects of AdS$_{2}$ quantum gravity and the Karch-Randall
  braneworld}}, \href{https://doi.org/10.1007/JHEP09(2022)024}{\emph{JHEP}
  {\bfseries 09} (2022) 024}
  [\href{https://arxiv.org/abs/2206.11277}{{\ttfamily 2206.11277}}].

\bibitem{Verheijden:2021yrb}
E.~Verheijden and E.~Verlinde, \emph{{From the BTZ black hole to JT gravity:
  geometrizing the island}},
  \href{https://doi.org/10.1007/JHEP11(2021)092}{\emph{JHEP} {\bfseries 11}
  (2021) 092} [\href{https://arxiv.org/abs/2102.00922}{{\ttfamily
  2102.00922}}].

\bibitem{KumarBasak:2021rrx}
J.~Kumar~Basak, D.~Basu, V.~Malvimat, H.~Parihar and G.~Sengupta, \emph{{Page
  curve for entanglement negativity through geometric evaporation}},
  \href{https://doi.org/10.21468/SciPostPhys.12.1.004}{\emph{SciPost Phys.}
  {\bfseries 12} (2022) 004}
  [\href{https://arxiv.org/abs/2106.12593}{{\ttfamily 2106.12593}}].

\bibitem{Anous:2022wqh}
T.~Anous, M.~Meineri, P.~Pelliconi and J.~Sonner, \emph{{Sailing past the End
  of the World and discovering the Island}},
  \href{https://doi.org/10.21468/SciPostPhys.13.3.075}{\emph{SciPost Phys.}
  {\bfseries 13} (2022) 075}
  [\href{https://arxiv.org/abs/2202.11718}{{\ttfamily 2202.11718}}].

\bibitem{Mumford}
E.~Sharon and D.~Mumford, \emph{{2D-Shape Analysis Using Conformal Mapping}},
  \href{https://doi.org/10.1007/s11263-006-6121-z}{\emph{Int J Comput Vision}
  {\bfseries 70} (2006) 55}.

\bibitem{Goto:2020wnk}
K.~Goto, T.~Hartman and A.~Tajdini, \emph{{Replica wormholes for an evaporating
  2D black hole}}, \href{https://doi.org/10.1007/JHEP04(2021)289}{\emph{JHEP}
  {\bfseries 04} (2021) 289}
  [\href{https://arxiv.org/abs/2011.09043}{{\ttfamily 2011.09043}}].

\bibitem{Fallows:2021sge}
S.~Fallows and S.~F. Ross, \emph{{Islands and mixed states in closed
  universes}}, \href{https://doi.org/10.1007/JHEP07(2021)022}{\emph{JHEP}
  {\bfseries 07} (2021) 022}
  [\href{https://arxiv.org/abs/2103.14364}{{\ttfamily 2103.14364}}].

\bibitem{Skenderis:1999nb}
K.~Skenderis and S.~N. Solodukhin, \emph{{Quantum effective action from the AdS
  / CFT correspondence}},
  \href{https://doi.org/10.1016/S0370-2693(99)01467-7}{\emph{Phys. Lett. B}
  {\bfseries 472} (2000) 316}
  [\href{https://arxiv.org/abs/hep-th/9910023}{{\ttfamily hep-th/9910023}}].

\bibitem{Dong:2020uxp}
X.~Dong, X.-L. Qi, Z.~Shangnan and Z.~Yang, \emph{{Effective entropy of quantum
  fields coupled with gravity}},
  \href{https://doi.org/10.1007/JHEP10(2020)052}{\emph{JHEP} {\bfseries 10}
  (2020) 052} [\href{https://arxiv.org/abs/2007.02987}{{\ttfamily
  2007.02987}}].

\bibitem{Ryu:2006bv}
S.~Ryu and T.~Takayanagi, \emph{{Holographic derivation of entanglement entropy
  from AdS/CFT}},
  \href{https://doi.org/10.1103/PhysRevLett.96.181602}{\emph{Phys. Rev. Lett.}
  {\bfseries 96} (2006) 181602}
  [\href{https://arxiv.org/abs/hep-th/0603001}{{\ttfamily hep-th/0603001}}].

\bibitem{Brown:1986nw}
J.~D. Brown and M.~Henneaux, \emph{{Central Charges in the Canonical
  Realization of Asymptotic Symmetries: An Example from Three-Dimensional
  Gravity}}, \href{https://doi.org/10.1007/BF01211590}{\emph{Commun. Math.
  Phys.} {\bfseries 104} (1986) 207}.

\bibitem{Banados:1998gg}
M.~Banados, \emph{{Three-dimensional quantum geometry and black holes}},
  \href{https://doi.org/10.1063/1.59661}{\emph{AIP Conf. Proc.} {\bfseries 484}
  (1999) 147} [\href{https://arxiv.org/abs/hep-th/9901148}{{\ttfamily
  hep-th/9901148}}].

\bibitem{Roberts:2012aq}
M.~M. Roberts, \emph{{Time evolution of entanglement entropy from a pulse}},
  \href{https://doi.org/10.1007/JHEP12(2012)027}{\emph{JHEP} {\bfseries 12}
  (2012) 027} [\href{https://arxiv.org/abs/1204.1982}{{\ttfamily 1204.1982}}].

\bibitem{Shimaji:2018czt}
T.~Shimaji, T.~Takayanagi and Z.~Wei, \emph{{Holographic Quantum Circuits from
  Splitting/Joining Local Quenches}},
  \href{https://doi.org/10.1007/JHEP03(2019)165}{\emph{JHEP} {\bfseries 03}
  (2019) 165} [\href{https://arxiv.org/abs/1812.01176}{{\ttfamily
  1812.01176}}].

\bibitem{Lewkowycz:2013nqa}
A.~Lewkowycz and J.~Maldacena, \emph{{Generalized gravitational entropy}},
  \href{https://doi.org/10.1007/JHEP08(2013)090}{\emph{JHEP} {\bfseries 08}
  (2013) 090} [\href{https://arxiv.org/abs/1304.4926}{{\ttfamily 1304.4926}}].

\bibitem{Dutta:2019gen}
S.~Dutta and T.~Faulkner, \emph{{A canonical purification for the entanglement
  wedge cross-section}},
  \href{https://doi.org/10.1007/JHEP03(2021)178}{\emph{JHEP} {\bfseries 03}
  (2021) 178} [\href{https://arxiv.org/abs/1905.00577}{{\ttfamily
  1905.00577}}].

\bibitem{Calabrese:2012nk}
P.~Calabrese, J.~Cardy and E.~Tonni, \emph{{Entanglement negativity in extended
  systems: A field theoretical approach}},
  \href{https://doi.org/10.1088/1742-5468/2013/02/P02008}{\emph{J. Stat. Mech.}
  {\bfseries 1302} (2013) P02008}
  [\href{https://arxiv.org/abs/1210.5359}{{\ttfamily 1210.5359}}].

\bibitem{Takayanagi:2017knl}
T.~Takayanagi and K.~Umemoto, \emph{{Entanglement of purification through
  holographic duality}},
  \href{https://doi.org/10.1038/s41567-018-0075-2}{\emph{Nature Phys.}
  {\bfseries 14} (2018) 573}
  [\href{https://arxiv.org/abs/1708.09393}{{\ttfamily 1708.09393}}].

\bibitem{Wen:2021qgx}
Q.~Wen, \emph{{Balanced Partial Entanglement and the Entanglement Wedge Cross
  Section}}, \href{https://doi.org/10.1007/JHEP04(2021)301}{\emph{JHEP}
  {\bfseries 04} (2021) 301}
  [\href{https://arxiv.org/abs/2103.00415}{{\ttfamily 2103.00415}}].

\bibitem{Baig:2022cnb}
S.~A. Baig and A.~Karch, \emph{{Double brane holographic model dual to 2d
  ICFTs}}, \href{https://doi.org/10.1007/JHEP10(2022)022}{\emph{JHEP}
  {\bfseries 10} (2022) 022}
  [\href{https://arxiv.org/abs/2206.01752}{{\ttfamily 2206.01752}}].

\end{thebibliography}\endgroup

\end{document}